\documentclass[showpacs
]{revtex4}
\usepackage{amssymb}
\usepackage{graphicx}

\newcommand{\bal}{\begin{align}}
\newcommand{\eal}{\end{align}}
\newcommand{\beq}{\begin{eqnarray}}
\newcommand{\eeq}{\end{eqnarray}}
\newcommand{\nneeq}{\nonumber \end{eqnarray}}

\newcommand{\nn}{\nonumber \\}
\newcommand{\es}{& = &}
\newcommand{\rs}{\, = \,}
\newcommand{\ps}{& + &}
\newcommand{\ms}{& - &}
\newcommand{\ts}{& \times &}
\newcommand{\nt}{\nn \ts}
\newcommand{\np}{\nn \ps}
\newcommand{\nm}{\nn \ms}

\newcommand{\cM}{ {\cal M} }
\newcommand{\cH}{ {\cal H} }

\newcommand{\cF}{ {\cal F} }
\newcommand{\cG}{ {\cal G} }

\newcommand{\cU}{ {\cal U} }
\newcommand{\cL}{ {\cal L} }

\begin{document}
\title{ Calculation of size for bound-state constituents 
\footnote{ These notes were written as a supporting material 
for the invited talk delivered by the author on 26th of May 
2014 at the conference {\it Light-Cone 2014, Theory and 
Experiment for Hadrons on the Light-Front}, NCSU, Raleigh, 
NC, USA, May 26-30, 2014.  } }
\author{       Stanis{\l}aw D. G{\l}azek        }
\affiliation{  Institute of Theoretical Physics,
               Faculty of Physics, 
               University of Warsaw             }
\date{         26 May, 2014                    }
\begin{abstract}

Elements are given of a calculation that
identifies the size of a proton in the
Schr\"odinger equation for lepton-proton bound
states, using the renormalization group procedure
for effective particles (RGPEP) in quantum field
theory, executed only up to the second order
of expansion in powers of the coupling constant.
Already in this crude approximation, the extraction 
of size of a proton from bound-state observables is
found to depend on the lepton mass, so that the
smaller the lepton mass the larger the proton size
extracted from the same observable bound-state
energy splitting. In comparison of Hydrogen and
muon-proton bound-state dynamics, the crude 
calculation suggests that the difference between 
extracted proton sizes in these two cases can be 
a few percent. Such values would match the order 
of magnitude of currently discussed proton-size 
differences in leptonic atoms. Calculations using 
the RGPEP of higher order than second are required 
for a precise interpretation of the energy splittings 
in terms of the proton size in the Schr\"odinger 
equation. Such calculations should resolve the 
conceptual discrepancy between two conditions: that 
the renormalization group scale required for high 
accuracy calculations based on the Schr\"odinger 
equation is much smaller than the proton mass (on 
the order of a root of the product of reduced and 
average masses of constituents) and that the energy 
splittings due to the physical proton size can be 
interpreted ignoring corrections due to the 
effective nature of constituents in the Schr\"odinger 
equation. 

\end{abstract} 
\maketitle

\vskip-4in
\hfill  IFT/14/02
\vskip4in

\section{ Introduction }

The approach suggested here for research on 
questions concerning the size of a proton in
lepton-proton bound states differs from several
approaches to bound-state dynamics that are 
available in the literature~\cite{BetheSalpeter, 
GellMannLow, BetheSalpeterBook, CaswellLepage, 
KinoshitaQED, Kinoshita1999, PachuckiMuon, 
PachuckiAtoms, Jones2, Kinoshita2012, Mohr2012, 
radius2013}. The suggested approach is not meant 
to easily achieve the accuracy that could rival 
advanced quantitative calculations. Instead, 
these notes address the conceptual issue of 
effective nature of constituents in the 
non-relativistic Schr\"odinger quantum 
mechanics. 

The constituents seen in bound-states are not 
considered here the same as the quanta of an 
underlying theory. They are instead treated as 
calculable, effective quanta in the renormalized 
theory. The required method of calculation is 
the renormalization group procedure for effective 
particles (RGPEP), whose recent summary can be 
found in Ref.~\cite{pRGPEP}. The effective 
particles and their interactions depend on the 
RGPEP scale parameter. The discussion of the 
proton size that follows accounts for the 
presence of such scale parameter in the 
two-body Schr\"odinger eigenvalue equation.

The effective nature of the two-body Schr\"odinger
approximation has an implication that the
lepton-proton interaction cannot be precisely
local. The non-locality is associated with the
finite value of the RGPEP scale parameter required
to justify the approximation of a bound state in
terms of just two constituents in the
Schr\"odinger equation. The range of the
non-locality is small in comparison to the
distances that characterize dominant effects in
the bound-state dynamics. The small non-locality
range can be estimated on the general grounds of
universality of the Schr\"odinger equation for
systems bound by electromagnetic interactions,
using scaling with the fine structure constant
$\alpha$. The point of this article is that the
non-locality matters in the interpretation of
corrections due to the physical proton radius 
with accuracy comparable to 1\%. 

The argument presented here is purely heuristic. 
It is based on a crude estimate of the coefficient, 
denoted here by $d_a$, in front of the proton radius 
parameter squared, $r_p^2$, in the ground-state 
lepton-proton binding energy correction due to the 
proton radius [cf. Eq.~(\ref{DeltaEda}) in 
Sec.~\ref{deltaVrp}],
\beq
\label{IntroDeltaEda}
\Delta E_r \es d_a \ { 2\pi \alpha \over 3} \
r_p^2 \ |\hat \psi(0)|^2  \ .
\eeq
The parameter $d_a$ must deviate from 1 in an 
effective theory. The size of deviation depends 
on the value of the number $a$ which says how 
much the RGPEP scale parameter $\lambda = 1/s$,
where $s$ has the interpretation of the size of
effective particles, differs from a special 
value selected on the basis of universality of 
electric charge and the Schr\"odinger quantum 
mechanics. One infers that
\beq
\label{l}
\lambda \es a \ \sqrt{ \mu M } \ ,
\eeq
where $\mu$ is the reduced and $M$ the average
mass of constituents in a two-body bound state.
With this choice one obtains the bound-state 
picture that universally scales with $\alpha$ if 
one uses the same $a$ in different systems. This 
means that one changes $\lambda$ when masses of 
the constituents change. On the other hand, when 
one uses the same $\lambda$ to discuss different 
systems in one and the same effective theory, one 
obtains different values of $a$ for different 
values of the constituent masses.

It follows from Eq.~(\ref{l}) that $a$ is reduced 
by the factor of about $\sqrt{ m_e/m_\mu} \sim 14$
when one goes from the muon-proton to electron-proton 
system without changing $\lambda$ in the effective 
theory. The associated change in the coefficient 
$d_a$ in Eq.~(\ref{IntroDeltaEda}) can reach the 
unexpectedly large values such as 8\%. The surprise
comes from the fact that the corrections result from 
a non-perturbative factor that, when formally expanded
in powers of $\alpha$, appears to correspond to 
corrections order $\alpha^4 \sim 10^{-9}$ to Rydberg.
The expansion is not valid, however, due to the 
ultraviolet divergences it illegitimately creates. 
On the other hand, without taking into account the 
non-perturbative result that $d_a \neq 1$ in 
Eq.~(\ref{IntroDeltaEda}), one might have an 
impression that the proton radius could be greater 
in the electron-proton system by about 4\% than in 
the muon-proton system. 

The possibility of obtaining an effect on the order
of a few percent in the proton radius extracted from 
one term in a theory of energy splittings does not
mean that the proton radius puzzle is explained. 
On the contrary, this finding provides evidence that 
the effective nature of Schr\"odinger picture for 
lepton-proton bound states must be brought under 
mathematical control as a function of the relevant 
scale parameter before one can arrive at firm conclusions. 
In exact calculations, observables cannot depend on 
the choice of scale for effective theory. But when 
one relies on approximations and one is not in 
possession of mathematical control over the effective 
theory dependence on the scale parameter, an artificial 
dependence of the proton radius on the system it is 
extracted from cannot be excluded. The RGPEP appears 
to be a candidate for attempts to remove this ambiguity 
and to compare the underlying theory with data including 
the proton radius effects.

These notes start with an introduction of the 
proton radius in Sec.~\ref{protonsize}. The 
effective nature of the Schr\"odinger equation 
is discussed in Sec.~\ref{effectivenature}, 
illustrated by a derivation in Sec.~\ref{derivation}. 
The effective size of the proton is considered in 
Sec.~\ref{size} and Sec.~\ref{C} concludes the 
notes. Appendix~\ref{cH} shows an example of 
construction of a canonical Hamiltonian. An 
outline of the RGPEP is given in
App.~\ref{ARGPEP}. Details of the effective 
lepton-proton interaction are discussed in 
App.~\ref{details} and a few details concerning
evaluation of $d_a$ are given in App.~\ref{Adf}.

\section{ Proton size in quantum mechanics }
\label{protonsize}

In the instant form (IF) of Hamiltonian 
dynamics~\cite{DiracFF}, the effective 
proton size in lepton-proton bound states can be 
introduced using the non-relativistic Schr\"odinger 
equation,
\beq
\label{Hpsi=Epsi}
{ \vec p^{\,2} \over 2\mu} \, \psi(\vec p\,)
+ \int {d^3 k \over (2\pi)^3 } \, 
V(\vec p, \vec k\,) \, \psi(\vec k\,)
\es 
-E_B \, \psi(\vec p\,) \ ,
\eeq
where the kernel $V(\vec p, \vec k\,)$ describes 
the interaction, $\vec k$ and $\vec p$ denote 
the relative momenta in the lepton-proton 
center-of-mass system before and after the 
interaction, $\mu$ is the reduced mass, and $E_B$ 
is the binding energy, which is defined as the 
difference between the sum of the constituent 
masses, expressed here by the average mass $M$,
\beq
\label{Mave1}
m_l + m_p  \es  2M \ ,
\eeq
and the bound-state mass $M_B$,
\beq
M_B \es 2M - E_B \ .
\eeq 
In the first approximation for lepton-proton bound 
states, $V(\vec p, \vec k\,)$ is equal to the 
spin-independent Coulomb potential for point-like 
charges, $V^{\rm pt}_C(\vec q\,)$, where $\vec q 
= \vec p - \vec k$ and 
\beq
V^{\rm pt}_C(\vec q \,) 
\es
- \ { 4 \pi \alpha \over \vec q^{\,2} } \, .
\eeq
In this approximation, the proton size can be
accounted for by the replacement of $V^{\rm pt}_C$ 
with
\beq
\label{VG}
V_C(\vec q \,) 
\es
V^{\rm pt}_C(\vec q \,) \ G_E(\vec q\,^2) \ ,
\eeq
where $G_E$ denotes the proton electric form 
factor, cf. Eq.~(2) in Ref.~\cite{radius2013}. 

The proton radius, denoted by $r_p$, enters 
in the interaction $V(\vec p, \vec k\,)$ in a 
known way because the momentum transfers $\vec 
q$ between constituents in the lepton-proton 
bound states are small in comparison with the 
inverse of the proton size and the proton 
form-factor dependence on such small momentum 
transfers is described by the formula 
\beq
\label{GE}
G_E(\vec q\,^2) = 1 - {1\over 6} \, r_p^2 \, \vec
q\,^2 + o(\vec q\,^2) \ .
%
\eeq
Thus, the interaction can be approximated by
\beq
\label{correction1}
V_C(\vec q \,) 
\es
V^{\rm pt}_C(\vec q \,) 
+ { 2\pi \alpha \over 3} \, r_p^2 \ .  
\eeq
In position variables,  
\beq
\label{correction2}
V_C(\vec r \,) 
\es
- { \alpha \over |\vec r\,|} 
+ { 2\pi \alpha \over 3} \, r_p^2 \ \delta^3(\vec r \,) \ .  
\eeq
Since the parameter $r_p$ is meant to describes a 
physical feature of the proton~\cite{radius2010e}, 
there is no obvious reason to expect that $r_p$ 
differs in the electron-proton and muon-proton 
bound states. 

However, splittings in the muon-proton bound-state 
spectra are measured with accuracy much better than 
the magnitude of corrections caused by the term 
\beq
\label{VCrpp}
\delta V(\vec p, \vec k\,) \es  { 2\pi \alpha \over 3} \, r_p^2 \ ,
\eeq
or, in its position representation, 
\beq
\label{VCrpr}
\delta V(\vec r\,)
\es
{ 2\pi \alpha \over 3} \, r_p^2 \ \delta^3(\vec r \,) \ .
\eeq
Differences between observed splittings in the 
spectra of electron-proton and muon-proton bound 
states~\cite{radius2010a} can be interpreted, 
including insight from Eqs.~(\ref{VCrpp}) and 
(\ref{VCrpr}), as resulting from $r_p$ being 
about 4\% greater in the electron-proton bound 
states than in the muon-proton bound 
states~\cite{radius2013}. While the available 
data suggest this interpretation, it is difficult 
to explain variation of the proton radius in theory. 
Therefore, the 4\% difference is called the proton 
radius puzzle. 

We suggest that a resolution of the proton radius 
puzzle may originate in details of the relationship 
between the Schr\"odinger equation, Eq.~(\ref{Hpsi=Epsi}), 
and quantum field theory (QFT). Namely, one needs to 
precisely define the steps through which the complex 
bound-state dynamics of relativistic QFT is reduced 
to a non-relativistic equation for the two particles 
that interact with each other through the instantaneous 
Coulomb potential, as a first approximation. Once 
this issue is clarified, the question then becomes 
if such precisely determined interactions can be
different in the electron-proton and muon-proton 
bound states and, if they can, if the relevant 
differences can be partly described by effectively 
changing the magnitude of parameter $r_p$ in 
$\delta V(\vec p, \vec k\,)$ by amounts comparable 
with the measured effect. We suggest that right 
answers to both of these questions may be positive. 
 
The procedure used here to define how the interaction
$V(\vec p, \vec k\,)$ can be calculated in QFT is 
called the renormalization group procedure for effective 
particles (RGPEP), which evolved from the similarity 
renormalization group procedure~\cite{GlazekWilson1993,
GlazekWilson1994} via introduction of the creation and
annihilation operator calculus~\cite{Glazek1994,Glazek1998}.
A succinct summary of a recent perturbative version of 
the RGPEP is available in Ref.~\cite{pRGPEP}. One can 
use the perturbative version because the relevant coupling 
constant, $\alpha \sim 1/137$, is small. Originally, 
the RGPEP was developed for calculating the effective 
Hamiltonians in QCD, where the coupling constant is 
large and one has to deal with confinement. The small 
value of $\alpha$ and the fact that leptons are not 
confined to protons, greatly simplify the lepton-proton
bound-state theory in comparison with QCD, but many of 
the steps in the procedure of deriving $V(\vec p, \vec k\,)$ 
for lepton-proton bound states are similar to the steps 
needed in QCD~\cite{Wilsonetal,QQ}. 

\section{ Effective nature of the Schr\"odinger equation }
\label{effectivenature}

Description of bound states in relativistic QFT
faces a conceptual difficulty which can be
identified in various ways. For example, if one
writes a bound-state equation using Feynman
diagrams~\cite{BetheSalpeter, GellMannLow}, one
needs the interaction kernel that properly
summarizes contributions of all relevant Green's
functions~\cite{DysonSchwinger}. If instead one
considers a Hamiltonian approach in some form of
dynamics~\cite{DiracFF}, one needs to account
in the Hamiltonian eigenvalue problem for
couplings among all relevant sectors in the
Fock-space. Both the Green's function and the
Hamiltonian approach generate divergences due to
integration to infinity over momenta of virtual
constituents. The conceptual difficulty is to
unambiguously limit the dynamics of an infinite 
and diverging set of amplitudes or wave functions 
to a manageable subset that can serve as a first 
step in a scheme of successive approximations.

For a bound state of a lepton and a proton, the
first-approximation picture is physically verified
to be the non-relativistic Schr\"odinger equation
for two particles interacting through the instantaneous 
Coulomb potential. If one wishes to connect perturbative 
diagrams with this non-perturbative picture, one needs 
a scheme of rules that allow one to decide which diagrams 
to include and how. If one wishes to use the Hamiltonian 
approach, one needs to account for the Fock sectors with 
additional photons and lepton-antilepton pairs, which all 
induce changes in the first-approximation Schr\"odinger 
eigenvalue problem. 

The Hamiltonian approach can be pursued using the
RGPEP. The physical proton radius is introduced in 
the initial theory and survives all steps of the 
procedure in which one reduces the complex QFT 
dynamics to the simple Schr\"odinger equation for 
two effective particles. Therefore, the RGPEP can 
be carried out as if the proton were point-like, 
which is simple to present, and then the proton 
radius effect can be inserted in the result. Finding 
a corresponding scheme in the diagrammatic approach 
would require separate research outside the Hamiltonian 
dynamics and such studies are not pursued here.

Consequently, the starting point of the RGPEP is 
here the canonical QFT Hamiltonian for leptons and 
protons coupled to photons that is described in 
Appendix~\ref{cH}. This Hamiltonian is regularized,  
supplied with counter-terms calculated using the 
RGPEP and eventually transformed into the scale-dependent 
effective Hamiltonian for which one can write an 
eigenvalue problem that resembles the 
first-approximation Schr\"odinger equation.
The physical proton radius inserted in the theory 
at the beginning re-surfaces there in the effective
two-body eigenvalue problem and illustrates how
the proton-radius puzzle can be addressed. 

Physical consequence of the scale dependence of 
the effective Hamiltonian is that the effective
constituents, the lepton and proton describable 
using the Schr\"odinger equation no longer 
interact through the pure Coulomb potential. 
Namely, the Coulomb potential is corrected by a 
scale-dependent form factor that limits the range 
of energy changes that the interaction can cause. 
This form factor introduces corrections that can 
be interpreted in terms of an apparent variation 
of the proton radius. Moreover, the magnitude of 
the variation can be changed by changing the 
scale of effective theory. The result for realistic
values of parameters is that the same physical 
proton radius shows up in the effective Schr\"odinger 
equations for muon-proton and lepton-proton bound 
states with different coefficients. The associated 
energy corrections follow from a formula that does 
not have a finite expansion in powers of the coupling 
constant around zero and the quantity one may 
interpret as a proton radius differs a bit from 
the physical proton radius, depending on the 
bound state one considers.

In summary, the lepton and proton that appear in
the non-relativistic Schr\"odinger equation for
electron-proton and muon-proton bound states are 
not the same as the bare quanta in QFT. Instead, 
they are the effective particles that are needed 
to represent the complex bound-state dynamics in 
QFT in terms of the universal Schr\"odinger picture 
of only two constituents and a potential. The price 
to pay for the simplification depends on the lepton 
mass and the corresponding corrections manifest 
themselves as if the proton radius depended on the 
lepton mass.

\section{ Derivation of the Schr\"odinger equation from QFT }
\label{derivation}

The bare regularized Hamiltonian 
that we start from is given in Appendix~\ref{cH}.
The counter-terms it requires are found using the 
RGPEP in the process of evaluating the Hamiltonian 
for effective particles of size $s$ (see below) and 
securing that its matrix elements in the basis states 
of small invariant mass do not depend on regularization. 
The effective Hamiltonian is evaluated here using 
second-order solutions to the RGPEP equations.
Subsequently, the eigenvalue equation for lepton-proton 
bound states in the whole effective-particle basis in 
the Fock space is artificially limited to just two 
sectors: the lepton-proton sector and lepton-proton-photon 
sector. The artificial limitation is legitimate because 
the effective interactions cannot change free invariant 
mass of interacting particles by much more than the 
inverse of their size and the ignored sectors do not 
contribute to the leading Coulomb potential in the 
lepton-proton sector. Reduction of the limited 
eigenvalue problem to the lepton-proton sector 
is carried out using a formal expansion in $\alpha$ 
and keeping only terms of order 1 and $\alpha$. The 
resulting FF Schr\"odinger equation for two effective 
particles is obtained in the form suitable for 
consideration of the proton-radius puzzle. The 
entire calculation resembles the RGPEP approach to 
physics of heavy quarkonia except that the handling 
of photons is much simpler than in the case of 
gluons because the effective photons are treated 
as massless~\cite{QQ}. The description that follows 
is limited to key points. 

The size of effective particles, $s$, is introduced 
in the RGPEP by defining the effective quantum fields, 
\beq
\label{qs}
\psi_s \es \cU_s \, \psi_0 \, \cU_s^\dagger \, ,
\eeq
where the transformation $\cU_s$ acts on the field 
$\psi_0$ which is the quantum field operator built 
from creation and annihilation operators for bare 
quanta of a local QFT. The canonical Hamiltonian 
density in Appendix~\ref{cH} is written using fields 
$\psi_0$. All bare creation and annihilation operators 
are commonly denoted by $q_0$. All creation and 
annihilation operators for effective particles of 
size $s$ are commonly denoted by $q_s$.  The field 
operators $\psi_s$ are built from operators denoted 
by $q_s$. 

The RGPEP starts with the equality
\beq
\label{cHt}
\cH_s(q_s) \es \cH_0(q_0) \, ,
\eeq
which says that the same dynamics is expressed in 
terms of different operators for different values 
of $s$. The Hamiltonian is assumed to be a polynomial 
in $q_s$ with coefficients $c_s$. For example, the 
canonical Hamiltonian in Appendix \ref{cH} only 
contains terms bilinear, trilinear and quadrilinear 
in $q_0$. For dimensional reasons, it is convenient 
to use the parameter $t = s^4$ instead of $s$. With 
the initial condition set at $t=0$, variation of the 
coefficients $c_t$ with $t$ is described by the RGPEP 
operator equation. Derivation of the FF Schr\"odinger 
equation requires solutions to the RGPEP equation 
for the operators order 1, $e$, and $e^2$, where $e$ 
denotes electric charges of fermions.

\subsection{ Initial condition }

To write down the required RGPEP solutions for $H_t$, 
we denote the initial condition at $t=0$ by $H_0 = 
\cH_0(q_0) = \hat P^- + CT$, where $\hat P^-$ is
the operator that is obtained from canonical 
Hamiltonian in the form of Eq.~(\ref{P234}) in 
Appendix \ref{cH} by replacing the fields $\psi_{n+}$ 
and $A^\perp$ in Eq.~(\ref{Pminus}) by the operators 
defined in Eqs.~(\ref{psir778}) and (\ref{Ar556}) 
and performing normal ordering. The initial condition 
includes the counter-terms $CT$ that are found using 
single fermion eigenvalue equations (see below). 

To order $e^2$, which is required for derivation
of the FF Schr\"odinger equation, we need only to 
consider
\beq
H_0 
\es 
H_{f \gamma} + \sum_{n = 1}^3 H_{n 0} \, ,
\eeq
where
\beq
H_{n0}
\es
H_{f n} + e_n Y_{n0} + e^2 \Sigma_{n 0} 
+ \sum_{l=1}^3 e_n e_l X_{n l 0} \, .
\eeq
The subscript $\gamma$ corresponds to photons 
and $n,l = 1, 2, 3$ correspond to electrons, muons 
and protons, respectively. The terms $H_f$ denote 
the free bilinear terms, $\Sigma$s stand for fermion
mass counter-terms (there is no need here for considering
the photon mass-squared counter-term), $Y$s stand for 
trilinear terms (there is no need for the counter-term 
of type $Y$ for the terms of lower order than $e^3$), 
and $X$s stand for quadrilinear terms in fields in 
Eq.~(\ref{Pminus}) in Appendix \ref{cH}. The
Counter-terms will be established below using the
effective eigenvalue equations for single fermions.

\subsection{ Solution for effective Hamiltonians }
\label{sH}

Solution obtained in Eqs.~(\ref{cHe}), (\ref{H1}) and 
(\ref{H2}) up to second order in a power series in 
$e$ for the terms in $H_t$ that are relevant to the 
FF Schr\"odinger lepton-proton bound-state eigenvalue 
problem, takes the form
\beq
\label{fullH}
H_t \es \cH_t(q_t) \rs H_{ft} + H_{It} \, ,
\eeq
where
\beq
\label{Hft}
H_{ft} 
\es 
H_{f\gamma}
+  
\sum_{n = 1}^3 H_{fn}  \, , \\
\label{HIt}
H_{It}
\es
f \sum_{n=1}^3 e_n Y_{n0} 
+ 
e^2 \sum_{n = 1}^3 
\left[ 
\Sigma_{n 0} 
+ 
\left( \cF Y_{n0} Y_{n0} \right)_\Sigma 
\right]
\np
f \sum_{n,l=1}^3 e_n e_l 
\left[ 
X_{n l 0} 
+ 
\left(\cF Y_{n0} Y_{l0} \right)_X 
\right] \, .
\eeq
All the terms in $H_t$ are polynomial functions of 
$q_t$. The free parts, i.e., all terms in $H_f$, differ
from $\cH_f$ only by replacement of the bare creation 
and annihilation operators, $q_0$, by the effective 
ones, $q_t$. The subscripts $\Sigma$ and $X$ indicate 
the extraction of operators proper for the term from a 
product of $Y$s. The form factors $f$ and $\cF$ are 
given in Eqs.~(\ref{f}), (\ref{cFold}) and (\ref{cFnew}).

\subsection{ Physical fermion states and mass counter-terms }

The eigenvalue equation for a fermion involves 
{\it a priori } infinitely many sectors no matter
what value of $t$ one uses. Although the smallest-mass
eigenstate with quantum numbers of a fermion $n$
is meant to represent a free physical particle
in empty space-time, the eigenstate is a combination 
of components with a virtual effective fermion, 
virtual effective fermion and photon, and virtual 
effective fermion with more photons than one and/or 
additional fermion-anti-fermion pairs. However, the 
larger the size $s$ of the effective particles, i.e., 
the larger $t=s^4$, the more restrictive the form 
factor $f$ in Eq.~(\ref{HIt}) for the effective 
interaction $H_{It}$. This means that the coupling 
of the virtual effective fermion to other sectors 
off shell decreases when $t$ increases. Therefore,
the spread of physical fermion states into the 
virtual Fock components decreases with increase
of $t$ and one can limit the physical-fermion 
eigenvalue problem to a few components when $t$
is sufficiently large. The value of $t$ required
for obtaining a universal Schr\"odinger picture
for lepton-proton bound states will be estimated
later on. 

Since the electromagnetic coupling constant 
$\alpha = e^2/(4 \pi \epsilon_0 \hbar c) \sim 
1/137$ is small, to fix the counter-term
$\Sigma_{n0}$ it is sufficient to solve the
eigenvalue equation for a physical fermion
of type $n$ with accuracy to the terms order
$\alpha$ in the formal expansion in powers 
of $e$. Thus, the physical-fermion state of
momentum $p$ and spin $s$ can be approximately 
represented in terms of effective particles of 
size $s$ by writing
\beq
|n\rangle 
\es
|1\rangle + |2\rangle \, ,
\eeq
where 1 refers to one effective fermion and 
2 to fermion-photon component at scale $t$.

The two-particle component can be eliminated 
from the eigenvalue equation using perturbation 
theory~\cite{Bloch,Wilson1971}. The reason is that 
the component $|2\rangle$ is generated from component 
$|1\rangle$ with factor $e_n Y$, which is formally 
of order $\sqrt{\alpha}$, and the probability of 
two effective constituents is of formal order 
$\alpha$. So, the eigenvalue problem 
\beq
H_t |n\rangle \es p^- |n\rangle \, ,
\eeq
in the approximation to just two sectors
is a pair of coupled equations (we drop $t$)
\beq
H_f       |2\rangle + f e_n Y_{n0}               |1\rangle \es p^- |2\rangle \, , \\
f e_n Y_{n0} |2\rangle + (H_f + e^2 \Sigma_{n0}) |1\rangle \es p^- |1\rangle \, , 
\eeq
and the effective Hamiltonian in sector
$|1\rangle$ has the form 
\beq
H_1 
\es 
H_f + e^2 \Sigma_{n0} + e^2 \left( \cF Y_{n0} Y_{n0} \right)_\Sigma 
+ 
e^2 \left( f Y_{n0}  \, {1 \over p_1^- - p_2^-} \, f
Y_{n0} \right)_\Sigma \, .
\eeq
The fermion eigenvalue has the form $p^- =
(p^\perp\,^2 + m_{n \, \rm phys}^2)/p^+$.

According to Appendix \ref{ARGPEP}
and using notation adopted in Eqs.~(\ref{cFold})
and (\ref{cFnew}), the eigenvalue condition takes 
the form (we drop $n$)
\beq
\label{fermion}
p_1^- 
+ 
e^2 \Sigma_0
+ 
e^2 
\sum_2 { p_{12} \, 12 + p_{12} \, 12 \over 12^2 + 21^2} \, 
\left[ 1 - e^{-t( 12^2 + 21^2)} \right] 
Y_{0 12} Y_{021}
+
e^2 \sum_2 f Y_{0 12}  \, {1 \over p_1^- - p_2^-} \, f Y_{0 21} 
\es p^- \, .
\eeq
This formula is written in detail to illustrate how the 
RGPEP calculation of a counter-term is carried out. The
formula can be considerably simplified by multiplication 
by $p^+$, canceling $p^\perp$ on both sides, expressing 
$\Sigma$ in terms of the mass-squared counter-term added 
to $m_n^2$, denoted by $e^2 \delta m_n^2$ (remembering 
that we drop $n$), observing that 
\beq
p_1^- - p_2^- \es ( m^2 - \cM_2^2 ) / p^+ 
\eeq
and $12 = - 21 = m^2 - \cM_2^2$, and using $f = \exp(-t \, 12^2)$. 
Namely, we have
\beq
m^2 + e^2 \delta m^2 
+ 
e^2 p^+ 
\sum_2 { 1 - f^2 \over p_1^- - p_2^-} \, 
Y_{0 12} Y_{021}
+
e^2 p^+ \sum_2 Y_{0 12}  \, {f^2 \over p_1^- - p_2^-} \, Y_{0 21} 
\es m_{\rm phys}^2\, ,
\eeq
which results in the expression for the
mass-squared counter-term,
\beq
\label{deltam2}
e^2 \delta m^2
\es m_{\rm phys}^2 - m^2 
-
e^2 p^+ \sum_2 { |Y_{0 12}|^2 \over p_1^- - p_2^-} \, ,
\eeq
as expected in second-order perturbation theory.
The same result is obtained for both of the RGPEP
generators in Eqs.~(\ref{cG1}) and (\ref{cG2})
because $ab=0$ in Eq.~(\ref{cFnew}). This result
is obtained in the no-cutoff limit on single 
fermion momenta $p^+$ and $p^\perp$ in quantum 
fields. An alternative way that leads to the same 
result is to use cutoffs on the relative $\perp$ 
momenta and fractions of $+$ momenta for quanta in 
the interaction vertices.

Thus, the fermion mass-squared counter-term 
$\Sigma_{n0}$ results in a rule: substitute 
$m_n^2 \to m_{n \, \rm phys}^2$ in $H_{f n}$ 
and ignore self-interactions. This rule works
in the lepton-proton bound state equations 
discussed in the next section. Therefore, one
can assume that the free Hamiltonian $H_f$
contains physical masses $m_{n \rm phys}^2$
instead of arbitrary $m_n^2$. This implies
that the fermion physical masses also appear
in the RGPEP form factors $f$ and $\cF$, {\it cf.}
Eqs. (\ref{f}), (\ref{cFold}) and (\ref{cFnew}). 

\subsection{ Lepton-proton bound-state equations }

In the discussion that follows we omit the subscript 
$t$, keeping in mind that the effective eigenvalue 
problem determines the bound-state wave functions 
for effective constituents of the RGPEP size $s = 
t^{1/4}$. We also omit the subscript $n$ for leptons, 
since one can focus on the electron-proton bound state 
knowing that the muon-proton bound state is described 
in a precisely analogous way by changing the electron 
mass to muon mass. 

The full-Fock-space bound-state eigenvalue problem, 
\beq
\label{HtB}
H |B\rangle \es E^- |B\rangle \, ,
\eeq
determines the eigenvalue 
\beq
E^- = (P^{\perp \, 2} + M^2_B)/P^+ \, ,
\eeq 
where $(P^+,P^\perp)$ denote kinematical components 
of the arbitrary total momentum and $M_B$ denotes 
the bound-state mass. In fact, $P^+$ and $P^\perp$
can be eliminated and the resulting boost-invariant 
eigenvalue equation determines the mass $M_B$, 
instead of the energy that depends on the bound-state
motion (see below). This boost-invariant eigenvalue 
equation will be shown below to reduce to the well-known 
two-body Schr\"odinger equation in the bound-state rest 
frame, with a tiny correction to the Coulomb potential 
that is related to the proton-radius puzzle.

Our reasoning is analogous to the one in Ref.~\cite{QQ} 
for heavy quarkonia, but we take advantage of the 
simplifications that occur thanks to the smallness 
of $\alpha$ and absence of confinement in the 
lepton-proton systems. For the RGPEP parameter 
$s$ much greater than the proton size, one can 
initially treat the proton as point-like and
subsequently account for its size by adding 
the required corrections.

The RGPEP vertex form factors in the effective Hamiltonian 
prevent effective particles from direct coupling to 
states with large virtuality and, since the coupling 
constant is small, the full eigenstate can be 
approximated by a superposition of just two sectors, 
\beq
\label{B}
|B\rangle 
\es
|2\rangle + |3\rangle \, ,
\eeq
where 2 refers to the effective lepton-proton and 3 
refers to the effective lepton-proton-photon states.
This approximation is sufficient for calculating 
the Coulomb potential in the effective electron-proton 
sector and observing the new feature of an effective 
theory that is relevant to the proton-radius puzzle. 

Using Eqs.~(\ref{fullH}), (\ref{Hft}), (\ref{HIt}) 
and (\ref{B}), one can write the eigenvalue 
Eq.~(\ref{HtB}) in the form 
\beq
(H_f + H_I) \, ( |2\rangle + |3\rangle )
\es 
 E^-        \, ( |2\rangle + |3\rangle ) \, .
\eeq
Since we work now in the subspace spanned by only 
two sectors made of effective particles, the 
eigenvalue problem is replaced by two coupled 
equations,
\beq
\left\{ H_f + e^2 
\sum_{n=1}^2
\left[ 
\Sigma_{n0} 
+ 
\left( \cF Y_{n0} Y_{n0} \right)_\Sigma 
\right]
-
f e^2 
\left[ 
X_0 
+ 
\left(\cF Y_{l0} Y_{p0} + \cF Y_{p0} Y_{l0}  \right)_X 
\right]
\right\} \, |2\rangle 
+ 
f e (Y_{l0} - Y_{p0}) |3\rangle 
\es 
E^- \, |2\rangle \, , \\
f e (Y_{l0} - Y_{p0}) |2\rangle + H_f |3\rangle )
\es 
E^- \, |3\rangle  \, .
\eeq
Following Refs.~\cite{Bloch,Wilson1971} and 
Eqs.~(56)-(58) in Ref.~\cite{QQ}, we obtain  
the reduced Hamiltonian in the electron-proton 
sector in the form
\beq
\label{HR}
H_2
\es
{1 \over \sqrt{ 1 + R^\dagger R} }
(1 + R^\dagger ) H (1+R)
{1 \over \sqrt{ 1 + R^\dagger R} }\, ,
\eeq
where the operator $R$ expresses the three-body
component in terms of the two-body component and 
to the first order in $e$ satisfies the equation
\beq
[R, H_f] \es feY_0 \, .
\eeq
The effective particle three-body component 
contributes to the two-body component eigenvalue 
equation through the operator $H_\gamma$ whose 
matrix elements between the effective two-body 
basis states $|i\rangle$ and $|j\rangle$ are
\beq
\langle i | H_\gamma | j \rangle
\es
{1 \over 2} 
\sum_3
\langle i |
f e (Y_{l0} - Y_{p0}) 
\left(
{1 \over p_i^- - p_3^-}
+
{1 \over p_j^- - p_3^-}
\right)
f e (Y_{l0} - Y_{p0}) 
| j \rangle \, .
\eeq
The denominators never cross zero because the 
invariant mass of a lepton-proton-photon state
is never smaller than the invariant mass of
the lepton-proton state in which the photon 
is created or which emerges as a result of
annihilation of the photon in the three-body 
sector. The denominators approach zero only 
when the photon momentum approaches zero, but
such long-wavelength photons decouple from 
neutral lepton-proton bound states. 

To order $\alpha$, the effective two-body 
Hamiltonian matrix elements including 
$H_{\gamma ij} = \langle i | H_\gamma | j \rangle$, 
are
\beq
\label{H2ij}
H_{2 ij} 
\es
H_{fij} + e^2 
\sum_{n=1}^2
\left[ 
\Sigma_{n0} 
+ 
\left( \cF Y_{n0} Y_{n0} \right)_\Sigma 
\right]_{ij}
-
f e^2 
\left[ 
X_0 
+ 
\left(\cF Y_{l0} Y_{p0} + \cF Y_{p0} Y_{l0}  \right)_X 
\right]_{ij} 
\np
{1 \over 2} 
\sum_3
f e (Y_{l0} - Y_{p0})_{i3} 
\left(
{1 \over p_i^- - p_3^-}
+
{1 \over p_j^- - p_3^-}
\right)
f e (Y_{l0} - Y_{p0})_{3j} 
\, .
\eeq
This Hamiltonian matrix is used to identify the 
concept of size for a proton in the Schr\"odinger
equation.

In terms of the matrix $H_{2ij}$, the lepton-proton 
bound-state eigenvalue problem reads
\beq
\label{se1}
\sum_j H_{2 ij} \psi_j \es E^- \psi_i \, ,
\eeq
where $\psi_i$ denotes the lepton-proton 
wave function in the component $|2\rangle$. 
Since the component $|2\rangle$ is made of
two fermions, the wave function $\psi_i$ 
for a lepton-proton bound state of total 
momentum $P$ and spin $S$, is denoted by 
$\psi^{P S}_{s_l s_p}(p_l,p_p)$, where the
subscript $l$ refers to lepton and $p$ 
to proton. Thus, the sum over $i$ in 
Eq.~(\ref{se1}) means summing over the 
constituents' spins and momenta. Using 
conventions introduced in Appendix~\ref{cH},
\beq
\label{psi}
|2\rangle
\es
\sum_{s_l s_p}\int_{p_l p_p} 
\ \psi^{P S}_{s_l s_p} (p_l,p_p) \
b^\dagger_{lt \, p_l s_l}
b^\dagger_{pt \, p_p s_p}
|0\rangle \ ,
\eeq
where $b^\dagger_{lt}$ and $b^\dagger_{pt}$ 
denote the creation operators for effective 
fermions corresponding to the RGPEP scale 
parameter $t=s^4$, see Eq.~(\ref{qs}) and 
Sec.~\ref{sH}. Normalization of the state 
$|2\rangle$ differs by a small amount from 
the normalization of the bound-state $|B\rangle$ 
due to the presence of component $|3\rangle$ 
in Eq.~(\ref{B}). In the leading approximation 
to be discussed below, the norm correction can 
be ignored, but it has to be accounted for in
a precise calculation.

\subsubsection{ Self-interaction terms }

The counter-term found in Eq.~(\ref{deltam2}) 
and the terms coming from the emission and 
absorption of effective photons in Eq.~(\ref{H2ij}), 
combine in the diagonal part of the effective 
two-body Hamiltonian, 
\beq
H_i \delta_{ij}
\es
H_{fij} + e^2 
\sum_{n=1}^2
\left[ 
\Sigma_{n0} 
+ 
\left( \cF Y_{n0} Y_{n0} \right)_\Sigma 
\right]_{ij}
+
{e^2 \over 2} 
\sum_3
\left( |f Y_{l0 \, i3}|^2 + |f Y_{p0 \, i3}|^2 \right)
\ {2 \over p_i^- - p_3^-} \ ,
\eeq
where 
\beq
H_i \es \sum_{n=l, p}  H_n 
\eeq
is a sum of terms for the lepton and proton. In 
the same fashion as in Eq.~(\ref{fermion}), 
but for two and three effective particle states
instead of for one and two, we have  
\beq
H_n \es p_n^- + e^2 \Sigma_{n0}
+ 
e^2 \sum_2 { 1 - f^2 \over p_n^- - p_{2/3}^- } \, 
|Y_{0 12}|^2 
+
e^2 \sum_3  \, { |f Y_{0 23}|^2 \over p_2^- - p_3^-} \, .
\eeq
The spectator fermion energy cancels in the 
denominator in the last term. Therefore, the
denominator is the same as the denominator in 
the third term, where $p_{2/3}$ denotes the 
momentum of the two interacting particles, 
out of the three including a spectator. These 
terms combine to
\beq
H_n \es p_n^- + e^2 \Sigma_{n0}
+ 
e^2 \sum_2 { |Y_{0 12}|^2 \over p_n^- - p_2^- } \,
.
\eeq
Using Eq.~(\ref{deltam2}) with $m^2 = m_{n \rm phys}^2$, 
one obtains the net result that the fermion counter-term 
cancels self-interactions: in order $\alpha$, the fermion 
masses in the bound-state eigenvalue equation are equal 
to their physical values, $m_{n \rm phys}^2$, as dictated 
by the spectrum of free physical particles described by
the same theory. Identification of effects such as the Lamb 
shift requires a calculation of the finite binding effects 
due to lepton-proton interaction in specified eigenstates,
rather than in the QFT effective Hamiltonian itself.

\subsubsection{ Lepton-proton interaction }

With counter-terms chosen above, the effective 
lepton-proton Hamiltonian matrix elements are
\beq
\label{H2ij1}
H_{2 ij} 
\es
H_{fij} 
-
f e^2 
\left[ 
X_0 
+ 
\left(\cF Y_{l0} Y_{p0} + \cF Y_{p0} Y_{l0}  \right)_X 
\right]_{ij} 
\nm
{e^2 \over 2} 
\sum_3
\left( f Y_{l0 i3} f Y_{p0 3j} + f Y_{p0 i3} f Y_{l0 3j} \right)
\left( {1 \over p_i^- - p_3^-} + {1 \over p_j^- - p_3^-} \right)
\, ,
\eeq
where the RGPEP factors $f$ and $\cF$ are given 
in Eqs. (\ref{f}), and (\ref{cFold}) or (\ref{cFnew}).
The sum over the three-body component in the effective
photon-exchange term amounts to the contraction of the 
effective photon annihilation and creation operators
and sum over photon polarizations. Therefore, using 
notation introduced in Eqs.~(\ref{f}) to (\ref{cFnew}), 
one can write the Hamiltonian matrix as
\beq
\label{H2ij2}
H_{2 ij} 
\es
H_{fij} 
-
e^2 W_{ij} \, ,
\eeq
where the interaction term is
\beq
W_{ij}
\es
f 
X_{0ij} 
+ 
f \cF \left( Y_{l0 i3} Y_{p0 3j} + Y_{p0 i3} Y_{l0 3j}\right) 
+
{ff \over 2} 
\left( Y_{l0 i3} Y_{p0 3j} + Y_{p0 i3} Y_{l0 3j} \right)
\left( {p_{i3} \over i3} + {p_{3j}\over j3} \right)
\, .
\eeq
Using the explicit formula for $\cF$, 
\beq
\cF_{i3j} 
\es
{ p_{i3} \, i3 + p_{j3} \, j3 \over i3^2 + 3j^2 - c \ ij^2} \, 
\left[ 1 - e^{-t( i3^2 + 3j^2 - c \ ij^2)} \right] \, ,
\eeq
with the coefficient $c$ equal 0 in the case of Eq.~(\ref{cFold})  
                           and 1 in the case of Eq.~(\ref{cFnew}), 
one arrives at
\beq
W_{ij}
\es
f_{ij} 
X_{0ij} 
+ 
\cF_{ij}
\left( Y_{l0 i3} Y_{p0 3j} + Y_{p0 i3} Y_{l0 3j}\right) 
\, ,
\eeq
where the complete RGPEP factor for the photon
exchange interaction is 
\beq
\cF_{ij}
\es
{ p_{i3} \, i3 + p_{j3} \, j3 \over i3^2 + 3j^2 - c \ ij^2} \, 
\left[ f_{ij} - f_{i3} f_{3j} f_{ij} f_{c \, ij}^{-1} \right] 
+
{f_{i3}f_{3j} \over 2} \left( {p_{i3} \over i3} + {p_{3j}\over j3} \right)
\, .
\eeq
The form factors $f$ are supplied here with subscripts 
to indicate the states which are used in evaluating 
their arguments. The factor $\cF_{ij}$ can be re-written 
as
\beq
\label{cFij}
\cF_{ij}
\es
f_{ij} 
\cF_{ij \, \rm int}
+
\cF_{ij \, \rm cor}
\, ,
\eeq
where 
\beq
\label{cFint}
\cF_{ij \, \rm int}
\es
{ p_{i3} \, i3 + p_{j3} \, j3 \over i3^2 + 3j^2 - c \ ij^2} \, 
\left[ 1 - f_{i3} f_{3j} f_{c \, ij}^{-1} \right] 
+
{f_{i3}f_{3j} \over 2} \left( {p_{i3} \over i3} +
{p_{3j}\over j3} \right) \, , \\
\label{cFcor}
\cF_{ij \, \rm cor}
\es
( 1 - f_{ij})
{f_{i3}f_{3j} \over 2} \left( {p_{i3} \over i3} + {p_{3j}\over j3} \right) \, .
\eeq
The labels ``int'' and ``cor'' correspond to 
the dominant interaction and a correction, 
respectively. 

The correction part $\cF_{ij \, \rm cor}$ is small
in the FF Schr\"odinger equation for lepton-proton 
bound states because $f_{ij}$ differs from 1 by 
terms on the order of relative lepton-proton 
momentum to fourth power, which is order $\alpha^4$, 
and the form factors $f_{i3}$ and $f_{3j}$ are small 
for the corresponding photon momentum. In brief, 
Eq.~(\ref{f}) shows that the fourth power comes from 
the square of a difference between the kinetic energies 
of leptons in relative motion with respect to proton 
before and after exchange of a virtual photon while 
the factor $f_{i3}f_{3j}$ secures exponentially fast 
fall-off with fourth power of the momentum transfer 
and the coefficient in the exponent is large, see 
App.~\ref{Aff}, Eqs.~(\ref{i3A}) and (\ref{j3A}). 
Thus, one can neglect the second term in Eq.~(\ref{cFij}) 
while making the crude estimates in this article. 
Precise calculations must include the second term.
 
With the dominant factor $\cF_{ij \, \rm int}$ in 
Eq.~(\ref{cFij}), the lepton-proton Hamiltonian 
matrix in Eq.~(\ref{H2ij2}) has the form 
\beq
\label{H2ij3}
H_{2ij} 
\es
H_{fij} 
-
e^2 f_{ij} V_{ij} \, ,
\eeq
where
\beq
\label{H2ij4}
V_{ij}
\es
X_{0 ij}
+
\cF_{ij \, \rm int}
\left( Y_{l0 i3} Y_{p0 3j} + Y_{p0 i3} Y_{l0 3j}\right) 
\, .
\eeq
Details of evaluation of $V_{ij}$ are summarized
in Appendix~\ref{details}. Using notation 
of Eqs.~(\ref{momentum1}) to (\ref{momentum2}),
illustrated in Fig.~\ref{Exchange}, and 
applying Eqs.~(\ref{AYijResult}), (\ref{XijA}) 
and (\ref{cFintA}), one obtains
\beq
V_{ij}
\es
\label{Vij}
\tilde \delta_{ij} \ \delta_{s_i  s_j} \delta_{r_i r_j} \
{  4 \sqrt{x(1-x)y(1-y)} \ (m_l + m_p)^2 
\over
(k^\perp - l^\perp)^2 + (m_l+ m_p)^2 z^2  } \ .
\eeq

\subsubsection{ The FF Schr\"odinger equation }
\label{sch}

Generally, the wave function $\psi^{P S}_{s_l s_p}(p_l,p_p)$
in Eq.~(\ref{psi}) can be written in the form 
\beq
\psi^{P S}_{s_l s_p}(p_l,p_p)
\es 
N \ P^+ \tilde \delta \
\psi_{s_l s_p}(x, k^\perp) \, ,
\eeq
where $N$ is the normalization factor, $\tilde \delta$
secures conservation of the total kinematical momentum 
of the bound state, and $\psi_{s_l s_p}(x, k^\perp)$ is a 
function of the FF boost-invariant relative momentum 
variables 
\beq
x \es p_l^+ /P^+ \, , \\
k^\perp \es (1-x) p_l^\perp - x p^\perp_p \, .
\eeq
The bound-state eigenvalue has the form 
\beq
E^- \es { P^{\perp \, 2} + M_B^2 \over P^+ } \, .
\eeq
Since the interaction in Eq.~(\ref{Vij}) does not
change spins, all four spin configurations of a 
lepton-proton system are described in the leading
approximation by one and the same wave function,
\beq
\psi_{s_l s_p}(x, k^\perp) 
\es
\psi(x, k^\perp) \, .
\eeq
In a precise calculation, spin splittings can be
treated in similar ways as in quarkonia. Projection 
of the FF Schr\"odinger Eq.~(\ref{se1}) with 
Hamiltonian of Eqs.~(\ref{H2ij3}) and (\ref{H2ij4}) 
on the lepton-proton basis states 
\beq
\label{basistekst}
|i\rangle
\es
b^\dagger_{lt \, l_i s_i}
b^\dagger_{pt \, p_i r_i}
|0\rangle \ ,
\eeq
yields in the notation of Eqs.~(\ref{momentum1}) 
to (\ref{momentum2}) and Fig.~\ref{Exchange}, 
the eigenvalue equation
\beq
\cM^2(x,k^\perp) \
\tilde \delta \ \psi(x, k^\perp) 
- e^2 
\int_{l_j p_j}  \, \tilde \delta_{ij} \
f \ \tilde V(x,k^\perp ; y, l^\perp) \
\tilde \delta \
\psi(y, l^\perp) 
\es
M_B^2 \ \tilde \delta \
\psi(x, k^\perp)  \, ,
\eeq
where 
\beq
\cM^2(x,k^\perp)
\es
{ k^{\perp \, 2} + m_e^2  \over   x}
+ 
{ k^{\perp \, 2} + m_p^2  \over 1-x} \, , \\
\label{fff}
f \es
e^{- t [\cM^2(x,k^\perp) - \cM^2(y,l^\perp)]^2} \, , \\
\tilde V(x,k^\perp ; y, l^\perp)
\es
{  4 \sqrt{x(1-x)y(1-y)} \ (m_l + m_p)^2 
\over
(k^\perp - l^\perp)^2 + (m_l+ m_p)^2 z^2  } \ .
\eeq
Integration over $l_j$ and $p_j$ allows one 
to factor out of the equation the momentum 
conservation $\delta$-function $\tilde \delta$ 
and one obtains the bound-state mass-eigenvalue 
problem in the form 
\beq
\label{FFSchroedinger1}
\cM^2(x,k^\perp) \ \psi(x, k^\perp) 
- e^2 
\int_{y \ l^\perp}  \  f  \ \tilde V(x,k^\perp ; y, l^\perp) \
\psi(y, l^\perp) 
\es
M_B^2 \ 
\psi(x, k^\perp)  \, ,
\eeq
where
\beq
\int_{y \ l^\perp} 
\es
\int 
{ dy \, d^2 l^\perp 
  \over 
  2y(1-y) (2\pi)^3 } \, .
\eeq
Both sides of the eigenvalue equation
can be divided by $2\sqrt{x(1-x)}$ to 
yield
\beq
\label{FFSchroedinger3}
\cM^2(x,k^\perp) \ \phi(x, k^\perp) 
- 2 e^2 
\int { dy \, d^2 l^\perp \over (2\pi)^3 } 
\  f  \ V(x,k^\perp ; y, l^\perp) \
\phi(y, l^\perp)
\es
M_B^2 \ \phi(x, k^\perp)  \, ,
\eeq
where
\beq
\label{Vraw}
V(x,k^\perp ; y, l^\perp)
\es
{  (m_l + m_p)^2 
\over
(k^\perp - l^\perp)^2 + (m_l+ m_p)^2 z^2  } \ ,
\eeq
and
\beq
\phi(x, k^\perp)
\es
{ \psi(x, k^\perp) \over 2\sqrt{x(1-x)} } \, .
\eeq
This is the raw form of the effective FF
Schr\"odinger equation for lepton-proton
bound states that one can derive from
QFT using the RGPEP. The new element 
in this equation that we focus on is 
the RGPEP form factor $f$.

\subsubsection{ The RGPEP form factor }
\label{ff}

The form factor $f$ of Eq.~(\ref{fff}) appears in 
Eq.~(\ref{FFSchroedinger3}) as the sole indicator of
the effective nature of the FF Schr\"odinger equation.
All other elements in the eigenvalue problem can be
derived in the canonical theory assuming that one 
can reduce the bound-state eigenvalue problem to the 
Fock sector of a bare proton and a bare lepton,
instead of the effective particles at the appropriate 
RGPEP scale. Even the lepton and proton mass terms
equal to their individually measureable values can 
be arrived at in the bound-state eigenvalue equation 
if one uses the approximation that a physical fermion 
is equal to a superposition of a bare fermion and a 
bare fermion-photon state, instead of a superposition
of the effective quanta. 

For large values of the RGPEP parameter $t = s^4 = 
1/\lambda^4$, which means for sufficiently small 
width parameter $\lambda$, the form factor strongly 
limits the range of values that the variable $z = x - 
y$ can take. It happens to be so because even a small 
change of the lepton momentum fraction from $y$ to 
$x$ results in a large value of the argument of the 
exponential function in $f$. Namely, in the form 
factor
\beq
f \es e^{- t (\Delta \cM^2)^2} \, , \\
\eeq
where 
\beq
\Delta \cM^2 
\es
\cM^2(x,k^\perp) - \cM^2(y,l^\perp) \ ,
\eeq
the argument of the exponential is $-t$ times
\beq
\Delta \cM^2 
\es
{ k^{\perp \, 2} + m_l^2  \over   x}
+ 
{ k^{\perp \, 2} + m_p^2  \over 1-x} 
-
{ l^{\perp \, 2} + m_l^2  \over   y}
- 
{ l^{\perp \, 2} + m_p^2  \over 1-y} 
\, .
\eeq
When $x$ differs from $y$, the mass-squared 
difference obtains a contribution from the 
mass terms
\beq
\label{DM2}
\Delta \cM^2_{\rm mass}
\es
(x-y) 
\left[ {m_p^2  \over (1-x)(1-y)} 
-
       {m_l^2  \over xy} \right] \, .
\eeq
This contribution is small if $z$ is small. 
But even a small value of $z$ produces a 
large value of $\Delta \cM^2_{\rm mass}$
if the coefficient of $z$, i.e., the square 
bracket in Eq.~(\ref{DM2}), is large. For 
the bracket to be small, $x$ and $y$ must 
be close to 
\beq
\beta \es { m_l \over m_p + m_l} \, , 
\eeq
cf. Eq.~(\ref{beta}). Writing $x = \beta + dx$
and $y = \beta + dy$, one obtains 
\beq
\Delta \cM^2_{\rm mass}
& \sim &
\left[ { (m_l + m_p)dx \over \sqrt{\beta(1-\beta)} } \right]^2 
-
\left[ { (m_l + m_p)dy \over \sqrt{\beta(1-\beta)} } \right]^2 
\, .
\eeq
This is a limit of small $dx$ and $dy$ in the
exact formula
\beq
\label{kz1}
\Delta \cM^2_{\rm mass}
\es
{(m_l + m_p)^2 \over \beta(1-\beta) } 
\left[ 
{ 
\sqrt{ \beta(1 - \beta) \over x(1 - x)} } \ 
(x - \beta) \right]^2 
-
{(m_l + m_p)^2 \over \beta(1-\beta) } 
\left[ 
{ 
\sqrt{ \beta(1 - \beta) \over y(1 - y)} } \ 
(y - \beta) \right]^2 
\, .
\eeq 
The latter form follows the definition of the 
effective constituent relative momentum introduced 
in Ref.~\cite{GlazekGluonBinding}, Eqs. (106) and 
(107). Namely,  
\beq
\label{kperp}
k^\perp_{\rm cons} 
\es 
\sqrt{ \beta(1 - \beta) \over x(1 - x) } \ 
k^\perp \ , \\
\label{kz}
k^z_{\rm cons} 
\es
\sqrt{ \beta(1 - \beta) \over x(1 - x) } \
(m_l + m_p) \ (x - \beta) \ .
\eeq
Using the three-dimensional variable $\vec k_{\rm cons}$, 
one obtains the free invariant mass of the lepton and 
proton system, with both particles assigned their physical 
masses, in the form
\beq
\label{M2cons}
\cM^2(x,k^\perp) 
\es (m_l+m_p)^2 + { \vec k_{\rm cons}^{\, 2} \over
\beta (1-\beta)} \, ,
\eeq
which implies 
\beq
\label{kz2}
\Delta \cM^2
\es
{ \vec k_{\rm cons}^2
- 
  \vec l_{\rm cons}^2
\over
\beta (1-\beta)} 
\eeq
in the RGPEP form factor $f$. The $z$-component 
part of this result is the content of Eq.~(\ref{kz1}).

Note that the definition of $k^\perp_{\rm cons}$ 
differs only by the constant factor $\sqrt{\beta
(1-\beta)}$ from the transverse momentum variable 
conjugated to the transverse relative position 
variable $\zeta^\perp$ in a quark-antiquark system 
in the light-front holography approach to hadronic 
physics~\cite{BTholography}, the latter being 
motivated by a correspondence between the light-front 
wave-function description of hadrons and AdS/CFT 
duality~\cite{Maldacena}. If QED is approached 
in the similar spirit, the two-body lepton-proton 
Schr\"odinger equation~\cite{Schroedinger} can be 
looked at as corresponding to QFT according to the 
Ehrenfest correspondence principle~\cite{Ehrenfest}.

It follows from Eqs.~(\ref{fff}) and (\ref{kz2})
for sizable values of the RGPEP parameter $t$ 
that the interaction term in Eq.~(\ref{FFSchroedinger3})
vanishes exponentially fast as a function of the 
difference between $|\vec k_{\rm cons}|$ and
$|\vec l_{\rm cons}|$. In addition, by writing 
$|\vec k_{\rm cons}|^2 - |\vec l_{\rm cons}|^2
= (|\vec k_{\rm cons}| + |\vec l_{\rm cons}|)
  (|\vec k_{\rm cons}| - |\vec l_{\rm cons}|)$
one can see that the form factor suppresses 
changes of the size of relative lepton-proton 
momentum exponentially stronger for large momenta 
than it does for small ones. At the same time, 
the interaction kernel $V$ of Eq.~(\ref{Vraw}) 
appears with a negative sign in Eq.~(\ref{FFSchroedinger3}) 
and the smaller the difference
\beq
\vec q
\es  
\vec k_{\rm cons} - \vec l_{\rm cons} 
\eeq
the stronger the interaction. Consequently, the 
interaction draws the relative-motion wave function 
of the bound lepton-proton system to configurations 
with momenta small in comparison with the constituents' 
masses (see below). This happens because the photon 
is massless and the denominator in $V$ may vanish 
for small $\vec q$.

In the absence of the RGPEP form factor $f$ in
Eq.~(\ref{FFSchroedinger3}), approximations to
QFT that are focused on the small values of $z$ 
and $k^\perp-l^\perp$, or $\vec q$, could not be
easily justified. A priori, the momenta could be 
very large because the numerator momentum-dependent 
spin factors, see App.~\ref{details}, could cause 
large regions of momentum to count, producing 
contributions that could compete in size with 
and even exceed the small-momentum contributions, 
especially when the momenta integrated over in the 
eigenvalue problem are allowed to be arbitrarily 
large and one uses the expansions in powers of 
momenta that are valid only for small momenta. 
Such long-range correlations in momentum space are 
eliminated by the RGPEP form factor, which is a 
characteristic feature of this method for deriving
effective theories. Precisely this feature of the
RGPEP makes it suitable for reducing a complex 
QFT dynamics to the effective-particle dynamics 
that is as simple as a two-body Schr\"odinger 
equation.

How large the changes of invariant masses of
effective particles can actually be due to 
their interactions is determined by the RGPEP 
scale $\lambda = 1/s$. It should be smaller 
than the proton mass to eliminate the interactions 
that can produce virtual proton-antiproton pairs. 
Such pairs could be created by photons in a local
theory, but pairs of composite baryons are not 
easily created by single photons.

On the other hand, the effects due to physical 
size of protons being order 1 fm may only be 
visible in the effective theory that includes 
contributions from momentum changes that are 
not negligible in comparison with the inverse fm, 
which means that $\lambda$ cannot be negligible 
in comparison with 200 MeV. To allow for the photon 
vacuum polarization caused by lepton-antilepton 
pairs, $\lambda$ should not be smaller than the 
lepton mass, since the vacuum polarization due 
to leptons is required in atomic physics, e.g., 
see~\cite{radius2013}. 

For the first approximation to effective theory
to match major features of the available QED picture, 
one may assume that $\lambda$ in the lepton-proton 
bound states is somewhere between the lepton and 
proton mass. Using these general arguments, one 
could propose that the average mass, 
\beq
\label{Mave}
M \es (m_l+m_p)/2 \ ,
\eeq 
is a candidate for the RGPEP $\lambda$ most useful
for discussing proton-size effects in atom-like
systems in simple terms. The choice of average mass 
makes sense also from the point of view of the 
assumption that the two-body forces in many-body 
systems do not significantly depend on the number 
of bodies, as if indeed the average mass were a 
suitable scale rather than a sum.

However, Eq.~(\ref{kz2}) for the RGPEP form 
factor $f$ includes the coefficient
\beq
{1 \over \beta(1-\beta)} \es 2 \ { M \over \mu } \ ,
\eeq
in front of the constituents' relative momentum
squared, where $\mu$ is the reduced mass. This 
coefficient depends on the constituents' masses 
in a more complex way than just through the average 
mass. At the same time, the Schr\"odinger equation 
with electromagnetic interactions is understood to 
be universally valid irrespectively of the values 
of reduced or average masses of the constituents. 

The verified universality of the Schr\"odinger
equation with the Coulomb interaction would demand 
that for a simple RGPEP derivation of the Schr\"odinger 
equation with electromagnetic interactions one 
uses $\lambda$ of such a value that the dependence 
of form factor $f$ on the constituent masses drops 
out. The resulting mass-independent quantum potential 
would match the Coulomb potential in the classical 
limit. 

A suitable choice for $\lambda$ is found on the basis 
of a well-known universal bound-state structure scaling 
with the coupling constant $\alpha$. The scaling is 
discussed below in Sec.~\ref{scaling} using a variational 
principle in the presence of the RGPEP form factor $f$.
The resulting effective potential matches exactly the 
Coulomb potential on-shell, which guarantees that it 
has the right form in the classical limit. The off-shell 
corrections one obtains to the Coulomb potential due to 
the RGPEP form factor $f$ are estimated in Sec.~\ref{size}. 
They turn out to matter in the interpretation of 
calculations that concern the effective proton radius.

\subsubsection{ Bound-state scaling with $\alpha$ }
\label{scaling}

The free invariant mass squared of effective 
constituents in Eq.~(\ref{FFSchroedinger3}) 
is shown in Eq.~(\ref{M2cons}) to be quadratic 
in $\vec k_{\rm cons}$. The interaction term 
in Eq.~(\ref{FFSchroedinger3}) involves the 
integral over $y$ or, equivalently, $l^z_{\rm 
cons} = z (m_l+m_p)$ that includes the mass 
factor taken out of the kernel $V$. The dynamically 
determined scale $p$ of momenta that characterize 
the ground-state, results from the minimum of 
the expectation value
\beq
E(p) \es { p^2 \over \beta (1-\beta)} - 
\tilde a \, 2e^2 \, (m_l+ m_p) \, p \ ,
\eeq
where $\tilde a$ is some dimensionless parameter 
(assuming $\hbar = c = 1$). The above estimate 
for $E(p)$ follows from the expectation value of 
a Hamiltonian (mass squared) of Eq.~(\ref{FFSchroedinger3}) 
in a trial state characterized by the momentum 
scale $p$ without consideration of any details 
of the wave function. Equating $\partial E/\partial
p$ to zero yields the estimate
\beq
p \es \tilde a \, e^2 \mu \, .
\eeq
This estimate is valid in the relativistic 
effective theory that includes the RGPEP
form factor $f$. The relativistic effects
that might yield a different scaling result
due to divergences, caused by spin factors 
for fermions and photons, are exponentially 
suppressed by $f$. 

Knowing that $p$ is proportional to $e^2 \mu$, or 
$\alpha \mu$, one can consider the limit of infinitesimal
$\alpha$ that facilitates the analysis based on the 
non-relativistic approximation. In this limit, momenta 
$\vec k_{\rm cons}$, $\vec l_{\rm cons}$ and $\vec q$ 
can be considered small in comparison with the reduced 
and average masses of the constituents, the former being
always smaller than the latter. The square roots in 
Eqs.~(\ref{kperp}) and (\ref{kz}) become 1 and 
the integral over $y$ changes to the integral over 
$l^z_{\rm cons}$ from $-\infty$ to $+\infty$. Once 
the wave function is written as
\beq
\phi(x, k^\perp) \es \psi(\vec k) \, , \\
\vec k & \equiv & \vec k_{\rm cons} \, ,
\eeq
etc., the effective FF lepton-proton bound-state
eigenvalue Eq.~(\ref{FFSchroedinger3}) takes the
form
\beq
\label{FFSchroedinger4}
\left[
(m_l+m_p)^2 + { \vec k^{\, 2} \over \beta (1-\beta)} 
\right]
\ \psi(\vec k \, ) 
- 2 \alpha (m_l+m_p)
\int { d^3 l \over (2\pi)^3 } \
{ f \ 4\pi \over (\vec k - \vec k)^2 } \
\psi(\vec l \, ) 
\es
M_B^2 \ \psi(\vec k \, )  \, ,
\eeq
where 
\beq
f \es 
\exp{ - \left[ 
      { \vec k^{\, 2} - \vec l^{\, 2} 
      \over 
      \beta (1-\beta)\lambda^2 } \right]^2 } \ .
\eeq
In distinction from the IF Schr\"odinger equation in 
non-relativistic quantum mechanics, the eigenvalue 
that comes out of the RGPEP effective Hamiltonian is 
the bound-state mass squared, instead of its energy.
The latter would depend on the frame of reference.
The RGPEP removes this difficulty. The origin of this 
relativistic result lies in the boost-invariance of
the FF of Hamiltonian dynamics. 

Taking advantage of the change of variables 
\beq
\vec k \es \alpha \mu \vec p \, , \\
\vec l \es \alpha \mu \vec p \,' \, ,
\eeq
to dimensionless variables $\vec p$ and $\vec p\,'$
that are expected to be of order 1 in the
ground-state solution, one arrives at
\beq
\label{FFSchroedinger5}
\left(
4M^2 + \alpha^2 2 M \mu \, \vec p^{\, 2} 
\right)
\ \psi(\vec p \, ) 
- 4 \alpha^2 \mu M
\int { d^3 p \,' \over (2\pi)^3 } \
{ f \ 4\pi \over (\vec p - \vec p\,')^2 } \
\psi(\vec p \,' ) 
\es
M_B^2 \ \psi(\vec p \, )  \, ,
\eeq
where $M = (m_l + m_p)/2$ and 
\beq
\label{f1}
f \es 
\exp{ - \left[ \alpha^2 \ { 2 \mu M \over \lambda^2} \
        \left( |\vec p\,|^2 - |\vec p\,'|^2 \right)
        \right]^2 } \ .
\eeq
This relativistic FF effective lepton-proton bound-state
equation can be reduced to its non-relativistic approximation 
using the smallness of $\alpha$. The reduction is required 
for comparison with the IF Schr\"odinger equation for 
the same system.

\subsubsection{ Non-relativistic limit of the FF Schr\"odinger equation }
\label{nrsch}

The non-relativistic approximation to
Eq.~(\ref{FFSchroedinger5}) is obtained by 
writing, cf. Eqs.~~(\ref{Mave1}) and (\ref{Mave}),
\beq
M_B \es 2M - E_B \, ,
\eeq 
and neglecting terms order $E_B^2$, which is 
equivalent of evaluating an approximate square 
root of Eq.~(\ref{FFSchroedinger5}). The 
dominant terms of order $M^2$ cancel out and 
the result is
\beq
\label{FFSchroedinger6}
\alpha^2 2 M \mu \, \vec p^{\, 2} 
\ \psi(\vec p \, ) 
- 4 \alpha^2 \mu M
\int { d^3 p \,' \over (2\pi)^3 } \
{ f \ 4\pi \over (\vec p - \vec p\,')^2 } \
\psi(\vec p \,' ) 
\es
- 4 M E_B \ \psi(\vec p \, )  \, ,
\eeq
which confirms that $E_B$ must tend to zero
as $\alpha^2$ when $\alpha$ tends to zero.
Division of both sides by $4 \alpha^2 \mu^2 M$ yields
\beq
\label{FFSchroedinger7}
{ \vec p^{\, 2} \over 2} 
\ \psi(\vec p \, ) 
- 
\int { d^3 p \,' \over (2\pi)^3 } \
{ f \ 4\pi \over (\vec p - \vec p\,')^2 } \
\psi(\vec p \,' ) 
\es
- {E_B  \over \alpha^2 \mu } \ \psi(\vec p \, )  \, .
\eeq
If the RGPEP form factor $f$ were set to 1
in this equation, the eigenvalues on its
right-hand side would be $-1/(2n^2)$ with
natural $n$. The wave functions would 
fall off as the fourth power of $|\vec p\,|$
for $|\vec p\,|$ much larger than 1. 

These results merely indicate that the 
equation with $f=1$ would match exactly 
the momentum-space version of the original 
Schr\"odinger equation~\cite{Schroedinger}
for an electron-proton bound state, written
in terms of the dimensionless variable $\vec p$
that describes the relative momentum of the
lepton with respect to the proton in units
of $\alpha \mu$. In fact, QED was built
on the basis of the Schr\"odinger equation 
maintaining its validity and there should be
no surprise in the RGPEP reproducing it in 
QFT of App.~\ref{cH}. 

It is clear from Eq.~(\ref{f1}) that setting
$f$ to 1 is justified if the RGPEP parameter
$\lambda$ is sufficiently large. Namely, for
\beq
\lambda^2 \gtrsim \mu M \ ,
\eeq
the form factor $f$ does not differ from 
1 over some considerable range of $|\vec p\,|$ 
below $\alpha^{-1}$. Since the wave functions
for $f=1$ fall off as $|\vec p\,|^{-4}$ for 
$|\vec p\,| \gg 1$, such limitation on 
$|\vec p\,|$ is not important numerically
in rough estimates. However, it will matter
in seeking high precision. Namely, by naive 
expansion of $f$ in powers of $\alpha$, one 
might expect a negative correction to Rydberg 
on the order of $\alpha^4 \sim 3 \cdot 10^{-9}$. 

On the other hand, $\lambda$ cannot be made
too large, as discussed in Sec.~\ref{ff}. 
The discussion suggests that the average 
mass of constituents, $M$, provides a 
reasonable estimate of the upper bound. 

There must exist an optimal value of $\lambda$ 
for the unquestionable physical accuracy of the 
Schr\"odinger equation in atomic physics to 
result from QFT already in the lowest-order 
RGPEP derivation described here. Such value 
of $\lambda$ must have the property that the 
corresponding Schr\"odinger equation follows
from the RGPEP irrespective of the masses of 
constituents, which is exemplified by the 
success of quantum mechanics in describing 
two-body systems greatly differing in masses 
of their constituents, such as positronium, 
muonium and the Hydrogen atom. This universality 
is also the basis of trust in the Schr\"odinger 
equation in description of systems such as deuteron 
and various quarkonia, despite that the underlying 
dynamics is quite different from electromagnetic. 
In those cases, the RGPEP indicates that the 
extended validity of the Schr\"odinger equation 
may still be based on the proper choice of 
effective constituents.

In the case of binding through the Coulomb 
potential, one can see in Eq.~(\ref{f1}) that 
the form factor $f$ will take a universal
shape irrespective of the constituents' masses
if $\lambda^2$ is proportional to the product
of the reduced mass $\mu$ and the averaged 
mass $M$,
\beq
\label{lambda}
\lambda^2 \es \, a^2  \ \mu \, M \, .
\eeq
The constant $a$ is not determined. There 
is no distinct reason known to the author 
for choosing $a$ that considerably differs 
from 1.

Varying constant $a$ means varying the RGPEP parameter 
$\lambda$. In exact RGPEP calculations, all observables, 
including the bound-state mass eigenvalues, are by 
construction entirely independent of $\lambda$. In 
contrast, changing the constant $a$ in $f$ in the 
approximate Eq.~(\ref{FFSchroedinger7}) does lead to 
minuscule changes in the interaction and hence also in 
the eigenvalues. To avoid changes in the eigenvalues, 
the coupling constant needs to vary with $a$, which is
explained in Ref.~\cite{optimization} in numerical detail 
using an exact RGPEP solution in a model. However, to 
correct for the effect of fourth power of $\alpha \sim 
1/137$ in the exponential in $f$, the required changes 
of $\alpha$ are very small, see next Section, and can 
be ignored in the first approximation, while a 
phenomenologically right value of $a$ is expected on 
the order of 1.

\section{ Effective size of the proton } 
\label{size}

Reinstating the Bohr momentum unit $\alpha \mu$
in Eq.~(\ref{FFSchroedinger7}) for comparison 
with Eq.~(\ref{Hpsi=Epsi}), one obtains
\beq
\label{Hpsi=EpsiRGPEP}
{ \vec p^{\,2} \over 2\mu} \, \psi(\vec p\,)
+ \int {d^3 k \over (2\pi)^3 } \, 
V_a(\vec p, \vec k\,) \, \psi(\vec k\,)
\es 
-E_B \, \psi(\vec p\,) \ ,
\eeq
where for the point-like proton the interaction
$V_a(\vec p, \vec k\,)$ has the form 
\beq
V_a^{\rm pt}(\vec p, \vec k\,)
\es 
f_a( p, k )
\
V_C^{\rm pt}(\vec q\,) \ , 
\eeq
and
\beq
f_a( p, k )
\es 
e^{ - 4( p^2 - k^2)^2/(a \, \mu)^4 } \ ,
\eeq
with the RGPEP parameter $a \sim 1$. According 
to Eq.~(\ref{VG}), the finite proton size can 
be included in the theory through multiplying 
the Coulomb potential for point-like proton, 
$V^{\rm pt}_C(\vec q \,)$, by the proton 
electric charge form factor $G_E(\vec q\,^2)$. 
In fact, the entire RGPEP calculation described 
here can be carried out with the proton form 
factors inserted into the Hamiltonian of QFT
in App.~\ref{cH}. As a result, the potential 
$V_a(\vec p, \vec k\,)$ obtains the form 
\beq
V_a(\vec p, \vec k\,)
\es 
f_a(p, k )
\
V_C^{\rm pt}(\vec q\,) 
\ 
G_E(\vec q\,^2) 
\eeq
and can be approximated by
\beq
\label{correction11}
V_a(\vec p, \vec k\,)
\es
f_a( p , k ) \
\left[
V^{\rm pt}_C(\vec q \,) 
+ { 2\pi \alpha \over 3} \, r_p^2 
\right]
\ .  
\eeq
The RGPEP form factor $f$ appears in the role of 
a regulator of the $\delta$-function potential in 
Eq.~(\ref{VCrpr}). The procedure thus removes an 
illusion that the non-relativistic Schr\"odinger 
equation describes dynamics of point-like charges. 
Instead, the equation applies to effective degrees 
of freedom whose RGPEP size scale corresponds to 
the inverse of the root of the product of their 
reduced and average masses.

In summary, the effective lepton and effective 
proton form a bound state due to the interaction
\beq
\label{correction12}
V_a(\vec p, \vec k\,)
\es
V^{\rm pt}_C(\vec q \,) 
+ 
\delta V_f
+ 
\delta V_{r_p}
\ ,
\eeq
where
\beq
\label{correctionVa}
\delta V_f
\es
\left[ f_a( p , k ) - 1 \right] \
V^{\rm pt}_C(\vec q \,) 
\eeq
and
\beq
\label{correctionVr}
\delta V_{r_p}
\es
f_a( p, k ) \ { 2\pi \alpha \over 3} \, r_p^2 
\ .  
\eeq
These corrections are discussed below separately 
one after another, focusing on estimates of
the proton radius.

\subsection{ Correction due to $\delta V_f$ }
\label{deltaVf}

The correction to binding energy, $E_B$, due 
to $\delta V_f$ in the theory that takes the 
proton size into account is the same as in 
the theory with a point-like proton. One may
think that its size can be estimated in 
perturbation theory by expanding the RGPEP 
form factor, 
\beq
f_a( p , k )
\es 
e^{ - 4( p^2 - k^2)^2/(a \mu)^4 } \\
& \sim &
1 - 4( p^2 - k^2)^2/(a \mu)^4 + O(\alpha^8) \ ,
\eeq
since momenta in the Schr\"odinger bound-state 
theory are on the order of $\alpha \mu$. The 
ground-state expectation value of the lowest-order 
correction to 1 thus appears to be 
\beq
\Delta E_f
\es
-
\int {d^3p \over (2\pi)^3}
\int {d^3k \over (2\pi)^3}
\  \psi(p) \
{ 4 ( p^2 - k^2)^2 \ 4\pi \alpha \over q^2 (a \mu)^4 } 
\ \psi(k) \ .
\eeq
However, the Schr\"odinger ground-state wave 
function $\psi$ falls off as fourth power of 
its argument. Therefore, the integral diverges 
and requires extra care to identify the role of 
$\ln \alpha$ in the answer. Instead, one can 
evaluate the correction numerically without 
expanding $f$,
\beq
\Delta E_f
\es
\int {d^3p \over (2\pi)^3}
\int {d^3k \over (2\pi)^3}
\  \psi(p) \
\left[ f_a( p , k ) - 1 \right] \
{ 4\pi \alpha \over q^2 } 
\ \psi(k) \ . 
\eeq
In terms of the dimensionless momentum variables
in units of $\alpha \mu$, the quantity to evaluate
is 
\beq
\label{DeltaEf}
\Delta E_f
\es
d_f \ { \mu \alpha^2 \over 2} \ ,
\eeq
where
\beq
\label{df}
d_f
\es
{
2 \int d^3p \int d^3p' \ 
\ { 1 \over (1 + p^2)^2 } \
\left[ e^{ - 4\alpha^4 ( p^2 - p'\,^2)^2/a^4 } - 1 \right] \
{4 \pi \over (\vec p - \vec p \,')^2 }  
\ { 1 \over (1 + p'\,^2)^2 } 
\over
(2\pi)^3 \int d^3p 
\ { 1 \over (1 + p^2)^4 } }
\ .
\eeq
For physical values of the parameters and for
$a=1$, one obtains $d_f \sim 2.4 \cdot 10^{-5}$.
Increasing $a$ to 2, yields $d_f$ nearly an order 
of magnitude smaller, while reducing $a$ to 1/2
produces $d_f$ nearly an order of magnitude larger.
Change of the ground-state wave function by the 
factor $\exp{[-4(p/\lambda)^4]}$ reduces the
correction by about 1/3 for $a=1/2$ and $a=1$, 
and by about a half for $a=2$. The corrections 
are thus on the order of $\alpha^3$ to $\alpha^2$
and as such are comparable with other corrections
of similar order, such as spin effects or the 
Lamb shift. It is clear that the effective
particle picture cannot be fully assessed without 
extensive calculations of new type for a whole
set of corrections that are already known to matter 
in other approaches.

One can compare the correction due to RGPEP form-factor 
to the vacuum-polarization correction due to electron-positron 
pairs in photon propagation in the electron-proton bound-state 
regime. The latter changes the coupling constant $\alpha(0)$ 
by the factor 
\beq
{ \alpha(Q^2) \over \alpha(0) }
& \sim &
1 + {\alpha(0) \over 15 \pi} \
Q^2/m_e^2 + O(\alpha^2) \ ,
\eeq
The RGPEP form factor $f_a$ can be crudely estimated by
\beq
f_a \es \exp{ [ - 4 (p+k)^2 (p-k)^2 /(a m_e)^4 ]} \\
& \sim & 
\exp{ [ - (2/a)^4 \alpha^2 Q^2 /m_e^2 ]} \ ,
\eeq
which corresponds to $\lambda = a \sqrt{\mu M}$
and replacements of $p+k$ by $2 \alpha m_e$ and 
$p-k$ by $Q$. The vacuum polarization introduces 
a positive effect of order 
\beq
v & \sim &  { \alpha \over 15 \pi } \ Q^2/m_e^2
\eeq
and the RGPEP form factor a negative effect of order
\beq
r & \sim & (2/a)^4 \alpha^2 \ Q^2/m_e^2 \ .          
\eeq
These two effects tend to cancel each other.
The ratio, 
\beq
r/v \es (2/a)^4 15 \pi \alpha 
\ \sim \ 0.34 \ (2/a)^4 \ ,
\eeq
says that the RGPEP form factor causes off-shell 
corrections to the Coulomb potential that are 
comparable with the corrections due to the 
vacuum polarization for $a \sim 1.5$. However, 
the RGPEP form factor deviates from 1 only 
off-energy-shell, while the vacuum polarization 
acts off- and on-energy-shell, which means that 
values of $a$ closer to 1 than 1.5 are more likely.

The above estimates show that an unambiguous
determination of corrections due to effective
nature of constituents in lepton-proton bound
states requires an RGPEP calculation carried out
with accuracy matching the contemporary QED
calculations~\cite{radius2013}. Such major
undertaking is far beyond the scope of this
article. The only statement one can make at this
point is that exact calculations of binding
energies must produce results that do not depend
on the RGPEP parameter $a$ and only the individual
corrections that come from different terms can
depend on $a$. 

Irrespective of the difficulty of precise
calculations, the correction $\Delta E_f$ does not
incorporate effects due to the proton size, which
is tiny on the atomic scale. Interpretation of
energy splittings in terms of the proton radius
depends instead on the correction caused by
$\delta V_{r_p}$.

\subsection{ Corrections due to $\delta V_{r_p}$ }
\label{deltaVrp}

Comparison of Eqs.~(\ref{VCrpp}) and (\ref{correctionVr})   
shows that the interpretation of observed energy splittings 
in terms of the proton radius should take into account that 
the Schr\"odinger equation provides a valid approximation 
to QFT if and only if one considers the constituents as 
effective particles of appropriate size scale $\lambda = 
a \sqrt{ \mu M}$. Therefore, the corrections due to physical 
proton radius should be interpreted using Eq.~(\ref{correctionVr}) 
rather than (\ref{VCrpp}). The results of measurement of 
relevant bound-state energy splittings should be compared 
with 
\beq
\label{DeltaEda}
\Delta E_r \es d_a \ { 2\pi \alpha \over 3} \ r_p^2 \ |\hat \psi(0)|^2  
\eeq
where $d_a$ is
\beq
d_a
\es
{ 1 \over |\hat \psi(0)|^2 } 
\int {d^3p \over (2\pi)^3}
\int {d^3k \over (2\pi)^3}
\  \psi(p) \
f_a(p , k )
\ \psi(k) \ ,
\eeq
assuming that the wave function is normalized to
1, in which case
\beq
\psi(\vec k\,) \es {N \over ( k^2 + \alpha^2 \mu^2)^2 } \ , \\ 
N \es 8 \sqrt{\pi} \ (\alpha \mu)^{5/2} \ , \\
|\hat \psi(0)|^2  
\es
{ (\alpha \mu)^3  \over \pi } \, .
\eeq
Thus, the correction due to the proton radius 
is
\beq
\label{DeltaEda1}
\Delta E_r \es 
d_a \ 
{ 4 \over 3} \ 
(\alpha \mu \ r_p)^2 \ 
{ \mu \alpha^2  \over 2} \ .
\eeq
Due to the ratio of the proton radius to the
Bohr radius, $\alpha m_e r_p \sim 10^{-5}/2$, 
this correction is order $10^{-10}/4$ times 
Rydberg in hydrogen atoms and about $(m_\mu/m_e)^3 
\sim 200^3 = 8 \ 10^6$ times larger in the 
muon-proton bound states.

Discussion of the size of $d_a$ can be carried
out using the momentum variables in units of 
$\alpha \mu$. Since $f_a \leq 1$, the result for 
$d_a$ is smaller than 1. Since $f_a$ does not
depend on the angles, one is left with the ratio
of two integrals
\beq
\label{daEq}
d_a
\es
{
\int_0^\infty dp \int_0^\infty dp' \ 
\ { p^2 \over (1 + p^2)^2 } \
e^{ - 4\alpha^4 ( p^2 - p'\,^2)^2/a^4 } 
\ { p'\,^2 \over (1 + p'\,^2)^2 } 
\over
\int_0^\infty dp \int_0^\infty dp' \ 
\ { p^2 \over (1 + p^2)^2 } \
  { p'\,^2 \over (1 + p'\,^2)^2 } 
}
\ .
\eeq
Since the exponential contains $(\alpha/a)^4$, one 
might think that $d_a$ differs from 1 by terms
order $(\alpha/a)^4 \sim 3 \cdot 10^{-9}/a^4$ . 

However, this estimate is false because the 
coefficient of $(\alpha/a)^4$ is badly divergent. 
The dominant part of the difference between 
$d_a$ and 1 cannot be calculated by the 
simplest expansion in powers of $\alpha$. 
Instead, one has to account for the limited 
range of the allowed invariant mass changes
in the theory of effective particles. The 
effective particles interact in a way that 
differs from the interactions of point-like
quanta in a non-perturbative way.

\begin{figure*}
\centering
\includegraphics[width=0.5\textwidth]{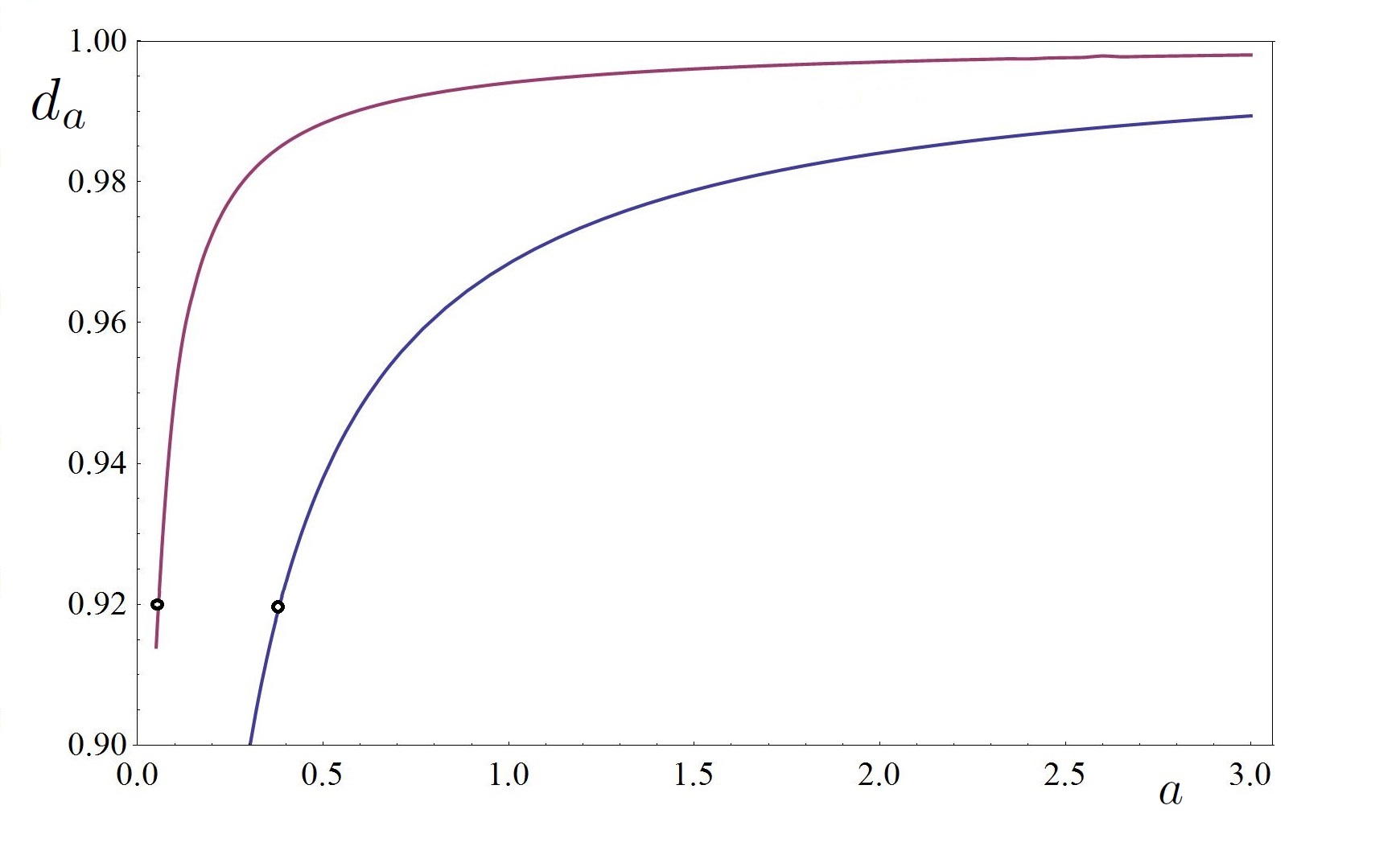}
\caption{ The lower curve shows the coefficient $d_a$ 
          in Eq.~(\ref{daEq}) as a function of the 
          RGPEP parameter $a$ in the region relevant 
          to phenomenology. The upper curve differs
          from the lower one by the rough estimate 
          of theoretical errors that is described in 
          the text. The dots on the curves indicate 
          the value $d_a = 0.92$ that might correspond 
          to a 4\% change in the proton radius extracted 
          from observed energy levels.}
\label{daFig}     
\end{figure*}

Fig.~\ref{daFig} shows the plot of $d_a$ as a 
function of the RGPEP parameter $a$ for 
$\alpha = 1/137.035999$.
The lower curve results from calculating $d_a$ from 
Eq.~(\ref{daEq}). The upper curve is obtained from 
a similar calculation in which only the ground-state 
wave function $(1+p^2)^{-2}$ is changed to $(1+p^2)^{-2}
\exp{[- 4\alpha^4 (p/a)^4]}$. This change is made
to mimic and thus estimate the effect on wave functions 
of the presence of the RGPEP form factor in the effective 
theory with $\lambda$ given in Eq.~(\ref{lambda}). If 
the wave functions were calculated in a precisely derived
effective theory that includes the RGPEP form factors,
the wave functions would not be the same as in the
ideal Schr\"odinger equation, Eq.~(\ref{Hpsi=Epsi})
with a local Coulomb potential for point-like particles.
Wave functions of eigenstates of a Hamiltonian including
the RGPEP form factor $f$ fall off faster for large 
momenta than the wave functions of eigenstates of
a Hamiltonian without $f$. This feature is modeled by 
the introduction of the exponential factor in the wave 
function only for orientation regarding the orders 
of magnitude. A rigorous estimate would require
an RGPEP calculation of the effective Hamiltonian 
including all terms in the expansion in powers of 
$\alpha$ that count in comparison of theory with 
data with current accuracy. As mentioned more than 
once before, such extensive research program is
far beyond the scope of this article, which merely
indicates a need for carrying out such program.

\subsection{ Interpretation of $d_a$ in terms of change in proton radius } 
\label{interpretation}

The effective Schr\"odinger Eq.~(\ref{Hpsi=EpsiRGPEP})
for lepton-proton bound states takes a universal 
scaling form of Eq.~(\ref{FFSchroedinger7}) when 
the value of the RGPEP parameter $\lambda$ is 
chosen differently for different leptons, see 
Eq.~(\ref{lambda}). This means that the standard
Schr\"odinger picture corresponds to choosing
\beq
\label{lambda1}
\lambda \es \, a  \ \sqrt{ \mu \, M } \ ,
\eeq
or
\beq
\label{lambdamu}
\lambda & \sim & \sqrt{\mu} \ .
\eeq
Since the electron and muon differ in masses
by the factor of about 200, the required choices
of $\lambda$ differ by the factor on the order
of 14. 

Variation of $\lambda$ required for maintaining 
one and the same scaling picture of the standard
Schr\"odinger quantum mechanics for different 
lepton-proton systems can be described using
the parameter $a$, which changes by the factor 
14 between electron-proton and muon-proton bound 
states. Fig.~\ref{daFig} shows that a change of 
such magnitude in $a$ can be correlated with a 
change in the coefficient $d_a$ in Eq.~(\ref{DeltaEda}) 
on the order of 8\%. Since the interpretation of 
the correction in terms of a change in the proton 
radius requires taking a square root of $d_a$, 
the accompanied variation in the extracted proton 
radius can be on the order of 4\%. Since $d_a$ 
is smaller for a lighter lepton, application of 
the same Schr\"odinger equation to both types of 
lepton-proton bound states will produce a greater 
proton radius for a lighter lepton.

If one uses the same effective theory for 
electron-proton and muon-proton bound states,
the form factor $f$ with one and the same 
value of $\lambda$ will fall off as a function 
of the relative electron-proton momentum in 
Hydrogen at the rate $\sqrt{m_\mu/m_e}$ times 
faster than as a function of the relative muon-proton 
momentum in a muon-proton bound state. The resulting 
reduction of range of the off-shellness in the 
interaction results in a reduction of the 
contribution to energy due to the proton-charge 
volume in the bound state and thus an increased 
estimate of the proton radius for a fixed value 
of the observed energy splitting.  

\section{ Conclusion } 
\label{C}

The RGPEP corrections in the Schr\"odinger equation 
due to the effective nature of bound-state constituents  
are discussed here in the leading approximation. The
corrections result solely from the form factors in 
the effective interactions. The form factors depend 
on the RGPEP scale parameter $s = 1/\lambda$ and make 
the Coulomb potential slightly non-local at short 
distances. The non-locality results from the upper 
bound order $\lambda$ on the changes of energy 
(actually, invariant mass) that can be caused by
the Coulomb interaction in the dynamics of effective
particles. 

More precisely, the upper bound on momentum changes 
in the RGPEP comes from an exponential function of 
fourth power of the ratio of momentum to the parameter 
$\lambda$. Since $\lambda$ needs to be on the order 
of masses for the effective theory to match the 
universal Schr\"odinger quantum mechanics with
electromagnetic interactions, the argument of the 
exponential function scales as $\alpha^4$, on top 
of the Schr\"odinger bound-state picture which 
yields binding energies that scale as $\alpha^2$. 
Despite the high power of $\alpha$, the RGPEP form 
factor can generate a noticeable correction in the 
extracted proton radius because it affects the
lepton-proton relative-motion wave function at the
origin.

The conceptual import of the RGPEP is that the 
bound-state constituents in the non-relativistic 
Schr\"odinger quantum mechanics are not point-like 
and they interact at short distances by a potential 
that slightly differs from the Coulomb potential 
for point-like charges. The effective nature of
constituents can be studied using the RGPEP in 
QFT.

Although the reasoning offered in this article is
focused on a specific term in the lepton-proton
bound-state dynamics, the RGPEP used in this
reasoning offers also access to corrections due 
to effective nature of particles in all areas
of physics where equations of the Schr\"odinger 
type apply. This means that in all such cases one 
is obliged to determine the scale of energy changes
that an effective interaction can cause and the
presence of such scale must be taken into account
in interpretation of precise comparisons between 
theory and experiment. 

\vskip.2in

{\bf Acknowledgment}

It is a pleasure to thank Krzysztof Pachucki for 
his comments on the author's ideas presented here.

\begin{appendix}

\section{ Canonical Hamiltonian } 
\label{cH}

The local action to consider is
\beq
S \es \int d^4x \, \cL
  \rs {1 \over 2} \int dx^+ dx^- d^2x^\perp \, \cL \, ,
\eeq
where 
\beq
\cL \es - {1 \over 4} F_{\mu \nu} F^{\mu \nu}
        + \sum_{n=1}^3 \bar \psi_n( i \partial \hspace{-5pt}/ 
        - e_n A \hspace{-5pt}/ - m_n) \psi \ ,
\eeq   
and subscripts $n=1, 2, 3$ refer to electrons, 
muons and protons, respectively. At this point,
proton is considered point-like and essentially 
of the same properties as leptons except for 
opposite charge and different mass. The 
corresponding canonical FF Hamiltonian in 
the gauge $A^+=0$ is~\cite{yan4}
\beq
P^- \es 
{1 \over 2} \int dx^- d^2x^\perp \, T^{+ \, -}
\ .
\eeq
We use the same convention for components of 
all tensors as in the case of Minkowski's 
space-time coordinates, for which
\beq
x^\pm   \es x^0 \pm x^3 \, , \\
x^\perp \es (x^1, x^2) \, .
\eeq
The co-ordinate $x^+$ plays the role of 
evolution parameter, or FF ``time,''
and $x^\perp$ and $x^-$ play the roles
of space co-ordinates in the front 
hyperplane in space-time.

The energy-momentum tensor density component 
$T^{+ \, -}$ is
\beq
\label{T+-3}
T^{+ \, -} 
\es  
A^\perp  \, (i\partial^\perp)^2 A^\perp
+ 
\sum_{n=1}^3 \psi_{n+}^\dagger 
\left[ ( i \partial^\perp \alpha^\perp + \beta m_n ) - e_n A^\perp \alpha^\perp \right] \, 
{2 \over i \partial^+} \, 
\left[ ( i \partial^\perp \alpha^\perp + \beta m_n ) - e_n A^\perp \alpha^\perp \right] \, 
\psi_{n+}
\np
\sum_{n=1}^3 2e_n \psi_{n+}^\dagger \psi_{+n} \, { 2 \over i\partial^+ } \, i\partial^\perp A^\perp
+ 
\sum_{n,l=1}^3 2e_n \psi_{n+}^\dagger \psi_{n+} \, 
{ 1 \over (i\partial^+)^2 } \, 2e_l
\psi_{l+}^\dagger \psi_{l+} \ ,
\eeq   
and for all fermion fields equally 
$\psi_\pm = \Lambda_\pm \psi $ and
$\Lambda_\pm = \gamma^0 \gamma^+/2$.
The Hamiltonian can be written as
\beq
\label{Pminus}
P^- 
\es  
\label{T+-4abbreviated}
\int dx^- d^2x^\perp \,
\left\{
{1 \over 2} A_\mu  \, \partial^{\perp \, 2} A^\mu
+ 
\sum_{n=1}^3
\left[
\bar \psi_n \gamma^+  
{ - \partial^{\perp \, 2} + m_n^2 \over 2 i \partial^+} \, 
\psi_n
\right.
\right.
\np
\left.
\left.
e_n \, \bar \psi_n A \hspace{-5pt}/ \, \psi_n
+
e_n^2 \, 
\bar \psi_n A \hspace{-5pt}/ \, 
         {\gamma^+ \over 2i \partial^+} \,
          A \hspace{-5pt}/ \, \psi_n
+
e_n \bar \psi_n \gamma^+ \psi_n \,          
{ 1 \over 2( i\partial^+)^2 } \,
\sum_{k=1}^3 \, e_k\,
\bar \psi_k \gamma^+ \psi_k 
\right]
\right\}
\, , \nn
\eeq  
where the dependent components of fields,
$A^-$ and $\psi_{n-}$, are solutions to the
constraint equations with all the electric 
charges $e_n$ set to zero. It is visible
that the Hamiltonian density contains terms
bilinear, trilinear and quadrilinear in the 
fields. Namely,
\beq
\label{Pminus2}
P^-_2
\es  
\int dx^- d^2x^\perp \,
{1 \over 2} A_\mu  \, \partial^{\perp \, 2} A^\mu
+ 
\int dx^- d^2x^\perp \,
\sum_{n=1}^3
\bar \psi_n \gamma^+  
{ - \partial^{\perp \, 2} + m_n^2 \over 2 i \partial^+} \, 
\psi_n
\, , \\
\label{Pminus3}
P^-_3 
\es  
\int dx^- d^2x^\perp \,
\sum_{n=1}^3
\,
e_n \, \bar \psi_n A \hspace{-5pt}/ \, \psi_n
\, , \\
\label{Pminus4}
P^-_4 
\es  
\int dx^- d^2x^\perp \,
\sum_{n=1}^3
\left[
e_n^2 \, 
\bar \psi_n A \hspace{-5pt}/ \, 
         {\gamma^+ \over 2i \partial^+} \,
          A \hspace{-5pt}/ \, \psi_n
+
e_n \bar \psi_n \gamma^+ \psi_n \,          
{ 1 \over 2( i\partial^+)^2 } \,
\sum_{k=1}^3 \, e_k\,
\bar \psi_k \gamma^+ \psi_k 
\right]
\, .
\eeq  

\subsection{ Quantization }
\label{quantization}

The quantum Hamiltonian is obtained by replacing
fields $A^\perp$ and $\psi_{n+}$ by field
operators $\hat A^\perp$ and $\hat \psi_{n+}$,
regulating the inverse powers of $i\partial^+$ in
the same way the field operators are regulated,
and normal ordering. Using the creation and
annihilation operators that are assumed to satisfy
the commutation relations
\beq
\label{bdcr}
\left\{ b_{ n p s}, b^\dagger_{ n' p' s'} \right\}
\es
\left\{ d_{ n p s}, d^\dagger_{ n' p' s'} \right\}
\rs
2p^+ (2\pi)^3 \delta^3(p - p') \, \delta_{s s'} \, \delta_{n n'}\, , \\
\label{acr}
\left[ a_{p s}, a^\dagger_{p' s'} \right]
\es
2p^+ (2\pi)^3 \delta^3(p - p') \, \delta_{s s'} \, ,
\eeq
with other relations being zero, respectively,
the field operators are written as (for 
our conventions concerning notation
for fermions, see~\cite{fermions})
\beq
\label{psir778}
\hat \psi_{n+}(x) 
\es 
\sum_s \int_p \ \Delta^{1/2}(p) 
\, \sqrt{p^+} \,
\left[  b_{nps} - d_{nps}^\dagger \, \sigma^1 \right] \, 
\left[
\begin{array}{c} \chi_s  \\
                 0       \end{array} 
\right] \, e^{-i c_p p \, x - \epsilon |x|} \, , \\
\label{Ar556}
\hat A^\perp(x) 
\es 
\sum_s \int_p \ \Delta^{1/2}(p) 
\left[  a_{ps}         \, \varepsilon_s^\perp      
      + a_{ps}^\dagger \, \varepsilon_s^{\perp *} \right]  
\, e^{-i c_p p \, x - \epsilon |x|}\, ,
\eeq
where 
\beq
\int_p \es \int {dp^+ \ d^2 p^\perp \over 2 p^+
(2\pi)^3 } \ \theta(p^+) \ ,
\eeq
and $\Delta(p)$ denotes the regularization 
function,
\beq
\Delta(p) \
\es
\Delta \left( |p^+|, |p^\perp| \right) \, .
\eeq
This function is required to tend to zero when momentum 
$|p^\perp|$ tends to infinity or $|p^+|$ tends to zero,
because divergences occur due to large $|p^\perp|$ and 
small $|p^+|$. Hence, the regularization requires two 
parameters. For example, if one used the regulator function 
of the form~\cite{Glazek1994} 
\beq
\label{Delta}
\Delta(p) \es 
\exp
\left( - \, { |p^\perp|^2 + \delta^2 \over
|p^+| \, \Delta } \right) \, ,
\eeq
the parameter $\Delta$ would limit $|p^\perp|$ from above 
and $\delta^2/\Delta$ would limit $|p^+|$ from below. In 
the no-cutoff limit, $\Delta$ tends to infinity and $\delta$ 
tends to zero. Other functions $\Delta(p)$ can be considered, 
especially such that factorize into the transverse and 
longitudinal regulating functions. The same function $\Delta$ 
is applied in regularization of the constraint equations. 
This regularization introduces factors $\Delta(p)$ in 
quadrilinear terms with $1/p^+$ and $1/p^{+ \, 2}$. 

The smooth exponential damping factors $e^{-\epsilon |x|}$ 
are introduced to eliminate boundary effects in a large 
quantization box in ``space'' directions of $x^\perp$ and
$x^-$. This means that one only focuses on the phenomena 
that fit well within the box of size $L \gg \epsilon^{-1}$. 
The principles of building the box are the same as in the 
formal scattering theory~\cite{GellMannGoldberger}. 

Spinors $\chi_s$ stand for the standard Pauli two component 
spinors and the photon polarization vectors are defined by 
writing $\varepsilon_s^\perp = (1,  i s)/\sqrt{2}$ with $s = 
\pm 1$ and the operators corresponding to $s=+1$ ($s=-1$) often 
labeled as 1(2) or $+$($-$). 

The coefficients $c_p$ in the Fourier-transform
exponentials are introduced for handling creation 
and annihilation operators in a generic operator 
calculus. We define $c_p$ by the rule that $c_p = 
1$ in a formula containing an annihilation operator 
with quantum numbers denoted by $p$, and $c_p = -1$ 
in a formula containing a creation operator with 
these quantum numbers. The generic factor of momentum 
conservation in interaction terms is denoted by 
\beq
\tilde \delta \es 2 (2\pi)^3 
\delta^3 \left( \sum_l c_{p_l} p_l \right) \, ,
\eeq
where $l$ runs from 1 to the number of fields in a term.

The quantum canonical Hamiltonian is a sum of terms that 
are bilinear, trilinear and quadrilinear in creation and 
annihilation operators. Namely, 
\beq
\label{P234}
\hat P^- 
\es 
\hat P^-_2
+ 
\hat P^-_3 
+ 
\hat P^-_4 \, ,
\eeq
where each term corresponds to its classical 
counterpart in Eqs.~(\ref{Pminus2}) to (\ref{Pminus4}). 
Explicit expressions for all terms in $\hat P^-$ 
are listed below in separate subsections.

\subsection{ Bilinear terms }

The bilinear terms are
\beq
\label{P2summary5}
\hat P^-_2 
\es
\sum_s \int_p \Delta(p) \
\left[
{p^{\perp \, 2} \over p^+} \,
a_{ps}^\dagger a_{ps} 
+
\sum_{n=1}^3 { p^{\perp  \, 2} + m_n^2  \over p^+}  \,
\left( b_{ps}^\dagger b_{ps} + d_{ps}^\dagger d_{ps} \right) 
\right] 
+ C_A + C_\psi \, , 
\eeq
where
\beq
C_A 
\es
{1 \over 2} 
\sum_s \int_p \ \Delta(p) \
2p^+(2\pi)^3 \delta^3(0) \, {p^{\perp \, 2} \over p^+} \, ,  \\
C_\psi
\es 
- \sum_s \int_p \ \Delta(p) \ 
2p^+(2\pi)^3 \delta^3(0) \, \sum_{n=1}^3 {p^{\perp \, 2} + m_n^2  \over p^+} 
\ .
\eeq
The constants $C_A$ and $C_\psi$ result from commuting
operators during normal ordering. As additive constants 
in the Hamiltonian, they could be ignored in quantum 
mechanics. However, one could include them in variational 
FF estimates of the vacuum energy if one wanted to recreate 
the vacuum effects known to cause problems in the IF of 
quantum field theory in $3+1$ dimensions~\cite{BartnikGlazek}.
Here, they are removed from the calculation.

\subsection{ Trilinear terms }

The trilinear terms are
\beq
\label{summary33}
\hat P_3^- 
\es 
\sum_{n=1}^3
\hat P_{3n} \, ,
\eeq
where 
\beq
\hat P_{3n}
\es
- e_n \sum_{s_1 s_2} \int_{p_1 p_2 q} 
\sqrt{ \Delta(p_1) \Delta(p_2) \Delta(q) } \ 
\tilde \delta \ \sqrt{ 2 p_1^+ p_2^+ } \ \hat Y_n \, , 
\eeq
and $\hat Y_n$ involves only operators and masses
for fermions number $n$, according to the same
pattern. Namely,
\beq
\label{Y}
\hat Y
\es
b^\dagger_{p_1 s_1}  b_{p_2 s_2} \, a_{q+} \, 
\left[
\left( {p_2 \over p_2^+} - {q \over q^+} \right) \, \delta_{s_1 1} \delta_{s_2 1}    
+
\left( {p_1 \over p_1^+} - {q \over q^+} \right) \,
\delta_{s_2 -1} \delta_{s_1 -1}
+
\left( { m \over p_2^+} - { m \over p_1^+} \right) \delta_{s_1 1} \delta_{s_2 -1} 
\right] 
\np
b^\dagger_{p_1 s_1}  b_{p_2 s_2} \, a_{q-}^\dagger \, 
\left[
\left( {p_2 \over p_2^+} - {q \over q^+} \right) \, \delta_{s_1 1} \delta_{s_2 1}    
+
\left( {p_1 \over p_1^+} - {q \over q^+} \right) \,
\delta_{s_2 -1} \delta_{s_1 -1}
+
\left( { m \over p_2^+} - { m \over p_1^+} \right) \delta_{s_1 1} \delta_{s_2 -1} 
\right] 
\np
b^\dagger_{p_1 s_1}  b_{p_2 s_2} \, a_{q-} \, 
\left[
\left( {p_2^* \over p_2^+ } - {q^* \over q^+} \right) \, 
\delta_{s_1 -1} \delta_{s_2 -1}
+
\left( {p_1^* \over p_1^+ } - {q^* \over q^+} \right) \, 
\delta_{s_2 1} \delta_{s_1 1}
+
\left( { m \over p_1^+} - {m \over p_2^+}\right) \, 
\delta_{s_2 1} \delta_{s_1 -1} 
\right] 
\np
b^\dagger_{p_1 s_1}  b_{p_2 s_2} \, a_{q+}^\dagger \, 
\left[
\left( {p_2^* \over p_2^+ } - {q^* \over q^+} \right) \, 
\delta_{s_1 -1} \delta_{s_2 -1}
+
\left( {p_1^* \over p_1^+ } - {q^* \over q^+} \right) \, 
\delta_{s_2 1} \delta_{s_1 1}
+
\left( { m \over p_1^+} - {m \over p_2^+}\right) \, 
\delta_{s_2 1} \delta_{s_1 -1} 
\right] 
\np
b^\dagger_{p_1 s_1}  d_{p_2 s_2}^\dagger \, a_{q+} \,
\left[
- \left( {p_2 \over p_2^+ } - {q \over q^+ } \right) \,      
\delta_{s_1 1} \delta_{s_2 -1} 
-
\left( {p_1 \over p_1^+} - {q \over q^+} \right) \,
\delta_{s_2 1} \delta_{s_1 -1}
+
\left( {m \, \over p_2^+} + { m \over p_1^+} \right) \,
\, \delta_{s_2 1} \delta_{s_1 1} 
\right]
\np
b^\dagger_{p_1 s_1}  d_{p_2 s_2}^\dagger \, a_{q-} \,
\left[
- \left( {p^*_2 \over p_2^+ } - {q^* \over q^+ } \right) \,      
\delta_{s_1 -1} \delta_{s_2 1} 
-
\left( {p_1^* \over p_1^+} - {q^* \over q^+} \right) \,
\delta_{s_2 -1} \delta_{s_1 1}
-
\left( {m \, \over p_2^+} + { m \over p_1^+} \right) \,
\, \delta_{s_2 -1} \delta_{s_1 -1} 
\right]
\np
d_{p_1 s_1} b_{p_2 s_2} \, a_{q+}^\dagger \,
\left[
- \left( {p^*_2 \over p_2^+ } - {q^* \over q^+ } \right) \,      
\delta_{s_1 1} \delta_{s_2 -1} 
-
\left( {p_1^* \over p_1^+} - {q^* \over q^+} \right) \,
\delta_{s_2 1} \delta_{s_1 -1}
+
\left( {m \, \over p_2^+} + { m \over p_1^+} \right) \,
\, \delta_{s_2 1} \delta_{s_1 1} 
\right]
\np
d_{p_1 s_1} b_{p_2 s_2} \, a_{q-}^\dagger \,
\left[
- \left( {p_2 \over p_2^+ } - {q \over q^+ } \right) \,      
\delta_{s_1 -1} \delta_{s_2 1} 
-
\left( {p_1 \over p_1^+} - {q \over q^+} \right) \,
\delta_{s_2 -1} \delta_{s_1 1}
-
\left( {m \, \over p_2^+} + { m \over p_1^+} \right) \,
\, \delta_{s_2 -1} \delta_{s_1 -1} 
\right]
\nm
d_{p_2 s_2}^\dagger d_{p_1 s_1} \, a_{q+} \, 
\left[
\left( {p_2 \over p_2^+ } - {q \over q^+} \right) \, 
\delta_{s_1 -1} \delta_{s_2 -1}
+
\left( {p_1 \over p_1^+ } - {q \over q^+} \right) \, 
\delta_{s_2 1} \delta_{s_1 1}
+
\left( { m \over p_1^+} - {m \over p_2^+}\right) \, 
\delta_{s_2 1} \delta_{s_1 -1} 
\right]
\nm
d_{p_2 s_2}^\dagger d_{p_1 s_1} \, a_{q-}^\dagger \, 
\left[
\left( {p_2 \over p_2^+ } - {q \over q^+} \right) \, 
\delta_{s_1 -1} \delta_{s_2 -1}
+
\left( {p_1 \over p_1^+ } - {q \over q^+} \right) \, 
\delta_{s_2 1} \delta_{s_1 1}
+
\left( { m \over p_1^+} - {m \over p_2^+}\right) \, 
\delta_{s_2 1} \delta_{s_1 -1} 
\right]
\nm
d_{p_2 s_2}^\dagger d_{p_1 s_1} \, a_{q-} \, 
\left[
\left( {p_2^* \over p_2^+} - {q^* \over q^+} \right) \, \delta_{s_1 1} \delta_{s_2 1}    
+
\left( {p_1^* \over p_1^+} - {q^* \over q^+} \right) \,
\delta_{s_2 -1} \delta_{s_1 -1}
+
\left( { m \over p_2^+} - { m \over p_1^+} \right) \delta_{s_1 1} \delta_{s_2 -1} 
\right] 
\nm
d_{p_2 s_2}^\dagger d_{p_1 s_1} \, a_{q+}^\dagger \, 
\left[
\left( {p_2^* \over p_2^+} - {q^* \over q^+} \right) \, \delta_{s_1 1} \delta_{s_2 1}    
+
\left( {p_1^* \over p_1^+} - {q^* \over q^+} \right) \,
\delta_{s_2 -1} \delta_{s_1 -1}
+
\left( { m \over p_2^+} - { m \over p_1^+} \right) \delta_{s_1 1} \delta_{s_2 -1} 
\right] 
\, ,
\eeq
where the subscript $n$ is omitted. The momentum 
variables $p_1$, $p_2$ and $q$ are complex numbers 
defined according to the rule
\beq
p \es p^1 + i p^2 \ .
\eeq
There are no terms resulting from commuting operators 
during normal ordering because of the momentum
conservation and presence of regularization factor 
$\Delta$ in the Fourier expansion of fields. 

\subsection{ Quadrilinear terms }

There are two kinds of quadrilinear terms.
One involves fermions and photons, denoted
by $\hat P^-_{4 \psi A}$, and the other 
one, analogous to the instantaneous Coulomb 
potential in the IF of dynamics, involves 
only fermions and is denoted by $\hat 
P^-_{4\psi \psi}$. These terms are 
described in two separate subsections.

\subsubsection{ Quadrilinear fermion-photon couplings $\hat P^-_{4 \psi A}$ }

The quadrilinear term $\hat P_{4\psi A}$ is 
a sum of terms for three kinds of fermions,
\beq
\label{PpsiA43}
\hat P_{4 \psi A}^- 
\es 
\sum_{n=1}^3 \hat P_{4 \psi_n A }^- \ ,
\eeq
where
\beq
\label{PpsiA33}
\hat P_{4 \psi_n A}^- 
\es 
e_n^2 \sum_{s_1 s_2 r_1 r_2} \int_{p_1 p_2 q_1 q_2} \
\sqrt{ \Delta(p_1) \Delta(p_2) \Delta(q_1) \Delta(q_2) } \
\tilde \delta \,
\sqrt{ p_1^+ p_2^+ } \ 
\, { 2 \Delta(c_{q_2} q_2 + c_{p_2} p_2) \over c_{q_2} q_2^+ + c_{p_2} p_2^+ } 
\ \hat Z_n 
\np
\sum_r \int_q \Delta(q) \
{ \delta m_{n \gamma}^2 \over q^+} \
a_{qr}^\dagger a_{qr}   
+
\sum_s \int_p \Delta(p) \
\left[
{ \delta m_n^2 \over p^+ }
\
b_{n p s}^\dagger b_{n p s}  
+
{ \delta m_{\bar n}^2 \over p^+ }
\
d_{n p s}^\dagger d_{n p s}
\right]
+
C_{\psi_n A} \ .
\eeq
Omitting the subscript $n$ in subscripts
of creation and annihilation operators 
for fermions,
\beq
\hat Z_n 
\es
\left[
\delta_{s_1 \, c_{q_1} r_1} \delta_{s_2 \, c_{q_1} r_1} 
\
b_{p_1 s_1}^\dagger  b_{p_2 s_2} 
-
\delta_{s_1 \, c_{q_1} r_1} \delta_{s_2 \, c_{q_2} r_2} 
\
b_{p_1 s_1}^\dagger d_{p_2 s_2}^\dagger 
\right.
\nm
\left.
\delta_{s_1 \, c_{q_2} r_2} \delta_{s_2 \, c_{q_1} r_1} 
\
d_{p_1 s_1} b_{p_2 s_2} 
-
\delta_{s_1 \, c_{q_2} r_2} \delta_{s_2 \, c_{q_2} r_2} 
\
d_{p_2 s_2}^\dagger d_{p_1 s_1}  
\right] 
\nt
\left[
a_{q_1 r_1} a_{q_2 r_2}      
+ 
a_{q_2 r_2}^\dagger a_{q_1 r_1}        
+
a_{q_1 r_1}^\dagger  a_{q_2 r_2}  
+
a_{q_1 r_1}^\dagger  a_{q_2 r_2}^\dagger 
\right] 
\
\delta_{ c_{q_1} r_1 \, - c_{q_2} r_2} 
\ .
\eeq 
The terms with only creation or only annihilation
operators do not actually contribute because of 
the conservation of $p^+$ and presence of the
regularization factors. Normal ordering proceeds 
through commuting operators and thus producing
the mass-like terms 
\beq
\label{PpsiA431}
\delta m_{n \gamma}^2 
\es
2 e_n^2
\int_p \
p^+ \ \Delta(p) \
\left[
{ \Delta(q - p) \over q^+ - p^+ } 
-
{ \Delta(q + p) \over q^+ + p^+ } 
\right] \ , \\
\delta m_n^2
\es 
2e_n^2
\int_q \ p^+ \
\Delta(q) \
{ \Delta( p - q ) \over p^+ - q^+ } 
\ , \\
\delta m_{\bar n}^2
\es
2e_n^2 
\int_q \ p^+ \
\Delta(q) \
{  \Delta( p + q ) \over p^+ + q^+ } \ , 
\eeq
and a number 
\beq
C_{\psi_n A}
\es
- 2 (2\pi)^3 \delta^3(0) \
4 e_n^2
\int_{pq}
\ \Delta(p) \Delta(q) \ p^+ \ { \Delta( p + q ) \over p^+ + q^+ } \ .
\eeq
All these terms are removed. They depend on the 
regularization function $\Delta(p)$ and ought to
be subtracted anyway. In the case of $\Delta(p)$
in Eq.~(\ref{Delta}) that correlates $\perp$ and
$+$ components of momentum, one may have to consider 
constants and operators of the type $e^2 \Delta 
A^\perp i\partial^+ A^\perp$ or $e^2 \Delta \bar 
\psi \gamma^+ \psi$ that do not obey regular FF 
power counting~\cite{Wilsonetal}.

Regarding protons, one should remember that they 
are not physically point-like in the sense that 
leptons are. The canonical terms in a local theory 
for protons is merely a method of book-keeping that 
applies only for the momentum transfers smaller than 
the inverse of their size.

\subsubsection{ Quadrilinear fermion couplings $\hat P_{4 \psi \psi}^-$ }

The fermion quadrilinear term $ \hat P^-_{4 \psi \psi} $ 
is the FF analog of the IF Coulomb term. It has the
form
\beq
\label{P-4-6}
\hat P^-_{4 \psi \psi} 
\es
\sum_{n=1}^3 \hat P^-_{4 \psi_n \psi} \, , 
\eeq
where
\beq
\hat P^-_{4 \psi_n \psi}
\es 
e_n \sum_{l = 1}^3 e_l 
\sum_{s_1 s_2 s_3 s_4} \int_{p_1 p_2 p_3 p_4} \, 
\sqrt{ \Delta(p_1)\Delta(p_2)\Delta(p_3)\Delta(p_4)} \
\tilde \delta \
\sqrt{ p_1^+ p_2^+ p_3^+ p_4^+} \
{ 2 \Delta( c_3 p_3 + c_4 p_4) 
\over (c_3 p_3^+ + c_4 p_4^+ )^2 } \
\hat X_{nl} \ .
\eeq
Using momentum conservation, properties of
the regularization factors, and performing 
normal ordering, one can write the terms 
that contribute to $\hat X_{nl}$
in the form
\beq
\hat X_{nl} 
\es  \hat X_{4nl} + \hat X_{2nl} + X_{0nl} \ .
\eeq
Genuine four-fermion interactions result
from
\beq
\label{X4}
\hat X_{4nl}
\es
- \delta_{s_1  s_2} 
  \delta_{s_3  s_4} \
  b_{n p_1 s_1}^\dagger  b_{l p_3 s_3}^\dagger 
  b_{n p_2 s_2}         b_{l p_4 s_4} 
\nm
  \delta_{s_1  s_2} 
  \delta_{s_3 -s_4} \
  b_{n p_1 s_1}^\dagger  b_{l p_3 s_3}^\dagger  
  d_{l p_4 s_4}^\dagger  b_{l n p_2 s_2} 
\nm
  \delta_{s_1  s_2} 
  \delta_{s_3 -s_4} \ 
  b_{n p_1 s_1}^\dagger  b_{n p_2 s_2} 
  d_{l p_3 s_3}         b_{l p_4 s_4}  
\np
  \delta_{s_1  s_2}
  \delta_{s_3  s_4} \ 
  b_{n p_1 s_1}^\dagger  d_{l p_4 s_4}^\dagger 
  b_{n p_2 s_2}         d_{l p_3 s_3}          
\nm
  \delta_{s_1 -s_2} 
  \delta_{s_3  s_4} \
  b_{n p_1 s_1}^\dagger  d_{n p_2 s_2}^\dagger 
  b_{l p_3 s_3}^\dagger  b_{l p_4 s_4} 
\np
  \delta_{s_1 -s_2} 
  \delta_{s_3 -s_4} \ 
  b_{n p_1 s_1}^\dagger  d_{n p_2 s_2}^\dagger 
  d_{l p_3 s_3}        b_{l p_4 s_4}  
\np
  \delta_{s_1 -s_2} 
  \delta_{s_3  s_4} \ 
  b_{n p_1 s_1}^\dagger  d_{n p_2 s_2}^\dagger 
  d_{l p_4 s_4}^\dagger  d_{l p_3 s_3}          
\nm
  \delta_{s_1 -s_2} 
  \delta_{s_3  s_4} \  
  b_{l p_3 s_3}^\dagger  d_{n p_1 s_1}          
  b_{n p_2 s_2}        b_{l p_4 s_4} 
\nm
  \delta_{s_1 -s_2} 
  \delta_{s_3 -s_4} \ 
  b_{l p_3 s_3}^\dagger  d_{l p_4 s_4}^\dagger
  b_{n p_2 s_2}         d_{n p_1 s_1}  
\np
  \delta_{s_1 -s_2} 
  \delta_{s_3  s_4} \ 
  d_{l p_4 s_4}^\dagger  d_{n p_1 s_1}          
  b_{n p_2 s_2}         d_{l p_3 s_3}          
\np
  \delta_{s_1  s_2} 
  \delta_{s_3  s_4} \ 
  d_{n p_2 s_2}^\dagger  b_{l p_3 s_3}^\dagger  
  d_{n p_1 s_1}         b_{l p_4 s_4} 
\np
  \delta_{s_1  s_2} 
  \delta_{s_3 -s_4} \ 
  d_{n p_2 s_2}^\dagger  b_{l p_3 s_3}^\dagger 
  d_{l p_4 s_4}^\dagger  d_{n p_1 s_1} 
\np
  \delta_{s_1  s_2} 
  \delta_{s_3 -s_4} \ 
  d_{n p_2 s_2}^\dagger  d_{n p_1 s_1} 
  d_{l p_3 s_3}         b_{l p_4 s_4}  
\nm
  \delta_{s_1  s_2} 
  \delta_{s_3  s_4} \ 
  d_{n p_2 s_2}^\dagger d_{l p_4 s_4}^\dagger  
  d_{n p_1 s_1}         d_{l p_3 s_3} 
\ .
\eeq
The mass-like terms in $\hat P_{4 \psi \psi}^-$ 
that result from commuting operators during normal 
ordering are due to $\hat X_{2nl}$. Namely, 
\beq
\hat P^-_{4 \psi \psi \, mass} 
\es 
\sum_s \int_p \Delta(p) \
{ \delta m_{\psi \psi}^2 \over p^+ }
\
\left( 
b_{p s}^\dagger  b_{p s} 
+ 
d_{p s}^\dagger  d_{p s}  
\right) \ , 
\eeq
where
\beq
\delta m_{\psi \psi}^2 
\es
2 e^2 \int_q 
\ p^+ q^+ \ 
\Delta(q) \
\left[ 
{ \Delta( p - q ) \over ( p^+ - q^+ )^2 } 
-
{ \Delta( p + q ) \over ( p^+ + q^+ )^2 } 
\right]
\, .
\eeq
Again, these depend on the regularization function
$\Delta$ and are removed. In the case of Eq.~(\ref{Delta})
one would have to consider similar terms as in 
$\hat P^-_{4 \psi A}$ but with additional
logarithms of $|\Delta i \partial^+|/\delta^2$.

The additive constant that one obtains in the 
canonical Hamiltonian by integrating
\beq
X_{0nl} 
\es
\delta_{nl} \ \left[
\delta_{s_1 -s_2} 
\delta_{s_3 -s_4} \ 
\delta_{23} \ \delta_{14} 
+
\delta_{s_1  s_2} 
\delta_{s_3  s_4} \ 
\delta_{12} \ \delta_{3 4} 
\right] \ ,
\eeq
can be ignored without any consequence in
the lepton-proton bound-state equation.

\section{ Outline of the RGPEP }
\label{ARGPEP}

The coefficients of powers of $q_t$ in $\cH_t(q_t)$
are found in the RGPEP using Eq.~(\ref{cHt}) and
calculating $\cH_t(q_0)$. Differentiation of 
\beq
\cH_t(q_0) \es \cU^\dagger_t \, \cH_0(q_0) \, \cU_t \, ,
\eeq
with respect to $t$ yields
\beq 
\label{ht1}
\cH'_t(q_0) \es
[ \cG_t(q_0) , \cH_t(q_0) ] 
\eeq 
with the generator $\cG_t = - \cU_t^\dagger \cU'_t$ and 
\beq
\label{Usolution}
\cU_t 
\es 
T \exp{ \left( - \int_0^t d\tau \, \cG_\tau
\right) } \, .
\eeq
$T$ denotes ordering in $\tau$. 
We consider the generator~\cite{Glazek1998,QQ}
\beq
\label{cG1}
\cG_t \es \left\{ (1-f^{-1})\cH_t \right\}_{\cH_f}
\eeq
and~\cite{pRGPEP}
\beq
\label{cG2}
\cG_t \es [ \cH_f, \cH_{Pt} ] \, .
\eeq
The operator $\cH_f$, called the free Hamiltonian, 
is the part of $\cH_0(q_0)$ that does not depend 
on the coupling constants, 
\beq
\label{cHf} 
\cH_f \es
\sum_i \, p_i^- \, q^\dagger_{0i} q_{0i} \, ,
\eeq 
where $i$ denotes particle species and $p_i^-$ 
is the free FF energy of a particle with mass 
$m_i$ and kinematical momentum components $p_i^+$ 
and $p_i^\perp$,
\beq
\label{pi-A}
p^-_i \es { p_i^{\perp \, 2} + m_i^2 \over p_i^+} \, .
\eeq
The curly bracket with subscript $\cH_f$ in
Eq.~(\ref{cG1}) means that by definition $\cG_t$ 
satisfies the equation $\left[ \cG_t, \cH_f \right] 
= (1-f^{-1})\cH_t$. The form factor $f$ depends on 
the difference between free invariant masses of 
the right (R) and left (L) sets of particles that
are involved in interaction in a Hamiltonian 
matrix element. Thus, R and L refer to the effective
particles that enter and emerge from the interaction.
The form factor is
\beq
\label{f}
f \es e^{-t(\cM_L^2 - \cM^2_R)^2} \, .
\eeq
The operator $\cH_{Pt}$ is defined for any polynomial 
$\cH_t$, 
\beq
\label{Hstructure} 
\cH_t(q_0) =
\sum_{n=2}^\infty \, 
\sum_{i_1, i_2, ..., i_n} \, c_t(i_1,...,i_n) \, \, q^\dagger_{0i_1}
\cdot \cdot \cdot q_{0i_n} \, ,
\eeq 
by multiplication of each and every term in it 
by a square of a total $+$ momentum involved
in a term,
\beq
\label{HPstructure} 
\cH_{Pt}(q_0) \es
\sum_{n=2}^\infty \, 
\sum_{i_1, i_2, ..., i_n} \, c_t(i_1,...,i_n) \, 
\left( {1 \over
2}\sum_{k=1}^n p_{i_k}^+ \right)^2 \, \, q^\dagger_{0i_1}
\cdot \cdot \cdot q_{0i_n} \, .
\eeq 
The multiplication secures that Hamiltonians 
$\cH_t$ possess 7 kinematical symmetries of 
the FF dynamics. The factor 1/2 is needed because
the sum includes both incoming and outgoing
particles that have the same total momentum.

Solutions to the RGPEP equation can be expanded
in powers of the charge $e$, 
\beq
\label{cHe}
\cH_t \es \cH_f + e \cH_t^{(1)} + e^2 \cH_t^{(2)}
+ ... \ .
\eeq
Up to order $e^2$ the bare charge and the renormalized
charge are the same and the terms of formal order $e$ 
and $e^2$ in the effective Hamiltonian read
\beq
\label{H1}
\cH_{t \, ab}^{(1)} \es f_{ab} \, \cH_{0 \, ab}^{(1)} \, , \\ 
\label{H2}
\cH_{t \, ab}^{(2)} \es f_{ab} 
\left[ 
\cH_{0 \, ab}^{(2)} 
+ 
\sum_x \cF_{axb} \cH_{0 \, ax}^{(1)} \cH_{0 \, xb}^{(1)}
\right] \, ,
\eeq
where, according to Eq.~(\ref{f})  with $L=a$ and $R=b$, 
$f_{ab} = \exp{(-t \, ab^2)}$. In case of the generator 
given in Eq.~(\ref{cG1}),
\beq
\label{cFold}
\cF_{axb} 
\es
{ p_{ax} \, ax + p_{bx} \, bx \over ax^2 + xb^2} \, 
\left[ 1 - e^{-t( ax^2 + xb^2)} \right] \, .
\eeq
In case of the generator given in 
Eq.~(\ref{cG2}),
\beq
\label{cFnew}
\cF_{axb} 
\es
{ p_{ax} \, ax + p_{bx} \, bx \over ax^2 + xb^2 -
ab^2} \, \left[ 1 - e^{-t( ax^2 + xb^2 -ab^2)} \right] \, .
\eeq
Both cases will lead to the same conclusion
concerning the proton radius in lepton-proton 
bound states. Extensive explanation of the 
symbols used in the above formulas can be found 
in~\cite{pRGPEP}. Symbols $a$, $b$ and $x$ denote 
the left, intermediate and right configurations 
of the effective particles that participate in 
the interaction, respectively. Symbols such as
$p_{ax}$ denote the total $p^+$ of particles
involved in the interaction, which is sandwiched
between states corresponding in this case to 
configurations $a$ and $x$. The invariant mass 
differences are denoted according to the rule 
$ ax = \cM^2_{ax} - \cM^2_{xa}$ and $\cM_{ax}$ 
denotes the invariant mass of the particles in 
$a$ that participate in the interaction that 
transforms $a$ into $x$. For example, in the 
left diagram in Fig.~\ref{Exchange}, $a = i$, 
$x = 3$, $b = j$, $p_{ax} = l^+(x,k^\perp)$, 
$p_{xb} = p^+(1-y, -l^\perp)$, $ax = m_l^2 - 
[q(z,q^\perp)+l(y,l^\perp)]^2$, $xb = [p(1-x,-k^\perp) +
q(z,q^\perp)]^2 - m_p^2$, and the minus components
of the momentum four-vectors for the lepton, $l$,
and proton, $p$, are calculated from the
mass-shell condition with masses $m_{l \, \rm phys}$
and $m_{p \, \rm phys}$, respectively.

Equations (\ref{H1}) and (\ref{H2}) illustrate the 
feature of $\cH_t(q_t)$ that its matrix elements in 
the effective-particle basis in the Fock space vanish 
exponentially fast as functions of the change of 
invariant mass due to interactions. The resulting 
band-width of the Hamiltonian matrix as measured in 
terms of the free invariant mass is denoted by 
$\lambda = 1/s$.

\section{ Details of the effective lepton-proton interaction }
\label{details}

\begin{figure*}
\centering
\includegraphics[width=0.5\textwidth]{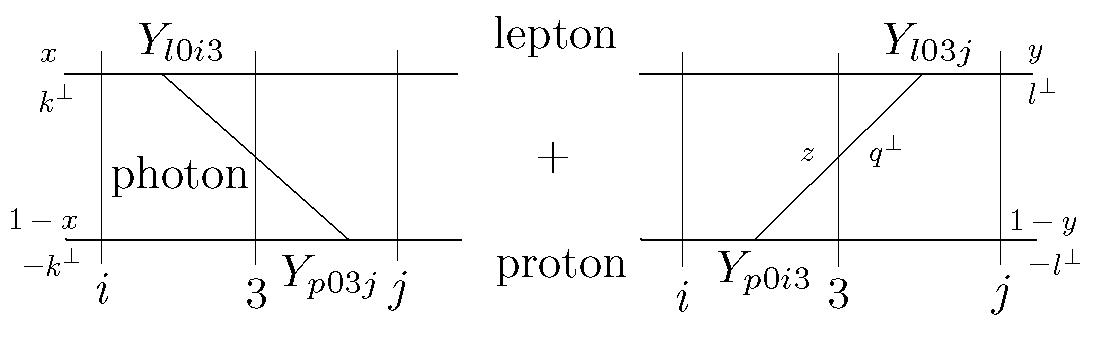}
\caption{ Exchange of the effective photon 
          between the effective lepton and 
          proton.}
\label{Exchange}     
\end{figure*}

Evaluation of the photon exchange in Eq.~(\ref{H2ij4}) 
is helped by Fig.~\ref{Exchange}. 
The operators $\hat Y_0(q_t)$ involve the fermion 
terms that are obtained from Eq.~(\ref{Y}) by 
putting the operators $q_t$ in place of $q_0$. 
Since anti-fermions do not contribute, the relevant 
operators $\hat Y_{n0}(q_t)$ with $n = l,p$ are 
\beq
\hat Y_{n0}
\es
- \sum_{s_1 s_2} \int_{p_1 p_2 q} 
\sqrt{ \Delta(p_1) \Delta(p_2) \Delta(q) } \ 
\sqrt{ 2 p_1^+ p_2^+ } \ \hat Y_n \, , 
\eeq
where $\hat Y_n$ involves only operators and masses
for fermions number $n$, in the common pattern exhibited 
by the first four terms in Eq.~(\ref{Y}). 

The FF instantaneous interaction term whose matrix
elements appear in Eq.~(\ref{H2ij4}) in addition to 
the photon exchange, $\hat X_0(q_t)$, contains the 
fermion-fermion terms obtained from Eq.~(\ref{X4}) 
by putting $q_t$ in place of $q_0$. The relevant part 
of $\hat X_0(q_t)$ for the matrix elements $X_{0ij}$ 
in Eq.~(\ref{H2ij4}) is 
\beq
\label{X0A}
\hat X_0
\es 
\sum_{n, l = 1}^2  
\sum_{s_1 s_2 s_3 s_4} \int_{p_1 p_2 p_3 p_4} \, 
\sqrt{ \Delta(p_1)\Delta(p_2)\Delta(p_3)\Delta(p_4)} \
\tilde \delta \
\sqrt{ p_1^+ p_2^+ p_3^+ p_4^+} \
{ 2 \Delta( p_4 - p_3) 
\over ( p_4^+ - p_3^+ )^2 } \
\hat X_{4nlA} \ ,
\eeq
where
\beq
\hat X_{4nlA}
\es
- (1-\delta_{nl}) \ \delta_{s_1  s_2} 
  \delta_{s_3  s_4} \
  b_{nt p_1 s_1}^\dagger  b_{lt p_3 s_3}^\dagger 
  b_{nt p_2 s_2}          b_{lt p_4 s_4} 
\ .
\eeq
No other operators than $Y$ and $X$ are needed in 
the evaluation of the lepton-proton interaction in 
the lepton-proton bound-state eigenvalue equation
up to $e^2$ in the formal series expansion in powers
of $e$ in the RGPEP. 

The matrix elements in Eq.~(\ref{H2ij4}) are
\beq
Y_{ij} 
\es
Y_{l0 i3} Y_{p0 3j} + Y_{p0 i3} Y_{l0 3j} 
\rs
\langle i|  \hat Y_{l0} \hat Y_{p0} + \hat Y_{p0} \hat Y_{l0} |j \rangle \, , \\
X_{ij}
\es
\langle i|  \hat X_0 |j \rangle  \, ,
\eeq
where the states $|i\rangle$ and $|j\rangle$ 
are created from vacuum by the effective lepton 
and proton creation operators. For example, 
\beq
\label{basisi}
|i\rangle
\es
b^\dagger_{lt \, ls}
b^\dagger_{pt \, pr}
|0\rangle \ .
\eeq
Evaluation of $Y_{ij}$ and $X_{ij}$ proceeds using
parameterization of momenta in Fig.~\ref{Exchange},
\beq
\label{momentum1}
l_i^+    \es    x  P^+             \, , 
\quad \quad
l_i^\perp \rs    x  P^\perp + k^\perp \, , \\
l_j^+    \es    y  P^+             \, , 
\quad \quad
l_j^\perp \rs    y  P^\perp + l^\perp \, , \\
p_i^+    \es (1-x) P^+             \, , 
\quad \quad
p_i^\perp \rs (1-x) P^\perp - k^\perp \, , \\
p_j^+    \es (1-y) P^+             \, , 
\quad \quad
p_j^\perp \rs (1-y) P^\perp - l^\perp \, , \\
q^+    \es      z  P^+             \, , 
\quad \quad
q^\perp \rs      z  P^\perp + \kappa^\perp \, ,
\label{momentum2}
\eeq
where $P^+$ and $P^\perp$ denote the total
momentum of the fermions. The total momentum 
is conserved by the interactions. 

\subsection{ Evaluation of $Y_{ij}$ }
\label{evYij}

In evaluating a matrix element of a product of two 
$\hat Y$s, one for a lepton and one for a proton,
one has to remember that one of the creation operators 
for one kind of fermions has to be commuted through 
the product of two operators of the other kind of 
fermions, with no net change of sign. The result for 
$Y_{ij}$ has the form
\beq
\label{AYij}
Y_{ij}
\es
\sqrt{ \Delta(p_i) \Delta(p_j) \Delta(l_i) \Delta(l_j)} 
\ \Delta(q) \ 
\sqrt{ 4 p_i^+ p_j^+  l_i^+ l_j^+} \
\tilde \delta \  \tilde Y_{ij}/q^+ \, ,
\eeq
where
\beq
\label{tildeYij}
\tilde Y_{ij}
\es
\theta(l_i^+ - l_j^+) 
\left[ c  (l_i, s_i,l_j, s_j) + m_l d(l_i,  s_i,l_j,  s_j)\right]          
\left[ c^*(p_i,-r_i,p_j,-r_j) - m_p d(p_i, -r_i,p_j, -r_j)\right]
\np 
\theta(l_i^+ - l_j^+) 
\left[ c^*(l_i,-s_i,l_j,-s_j) - m_l d(l_i, -s_i,l_j, -s_j)\right]
\left[ c  (p_i, r_i,p_j, r_j) + m_p d(p_i,  r_i,p_j,  r_j)\right]
\np
\theta(l_j^+ - l_i^+) 
\left[ c  (p_i, r_i,p_j, r_j) + m_p d(p_i,  r_i,p_j,  r_j)\right] 
\left[ c^*(l_i,-s_i,l_j,-s_j) - m_l d(l_i, -s_i,l_j, -s_j)\right]
\np 
\theta(l_j^+ - l_i^+) 
\left[ c^*(p_i,-r_i,p_j,-r_j) - m_p d(p_i, -r_i,p_j, -r_j)\right]
\left[ c  (l_i, s_i,l_j, s_j) + m_l d(l_i,  s_i,l_j,  s_j)\right]
\\
\es
\theta(l_i^+ - l_j^+) \ Y_+ + \theta(l_j^+ - l_i^+) \ Y_-
\, ,
\label{tildeYij1}
\eeq
and
\beq
c(l_i,s_i,l_j,s_j)
\es 
\left[
\left( {l_j \over l_j^+} - {q \over q^+} \right) \, \delta_{s_i 1} \delta_{s_j 1}    
+
\left( {l_i \over l_i^+} - {q \over q^+} \right) \, \delta_{s_j -1} \delta_{s_i -1}
\right] \, , \\
d(l_i, s_i,l_j, s_j)
\es
\left( {1 \over l_j^+} - {1 \over l_i^+} \right) \delta_{s_i 1} \delta_{s_j -1} 
\, , \\
c(p_i,r_i,p_j,r_j)
\es 
\left[
\left( {p_j \over p_j^+} - {q \over q^+} \right) \, \delta_{r_i  1} \delta_{r_j 1}    
+
\left( {p_i \over p_i^+} - {q \over q^+} \right) \, \delta_{r_j -1} \delta_{r_i -1}
\right] \, , \\
d(p_i, r_i,p_j, r_j)
\es
\left( {1 \over p_j^+} - {1 \over p_i^+} \right) \delta_{r_i 1} \delta_{r_j -1} 
\, .
\eeq
To explain how the calculation proceeds, 
it is enough to show details for $Y_+$,
which is
\beq
P^{+\,2} \ Y_+
\es\
\delta_{s_i  1}  \delta_{s_j  1} \ \delta_{r_i -1} \delta_{r_j -1}
\
\left[ 
\left( { l   \over   y} - {\kappa \over z} \right) 
\left( {-l^* \over 1 - y} - {\kappa^* \over z} \right) 
+
\left( {k^* \over x} - {\kappa^* \over z} \right) 
\left( {-k \over 1 - x} - {\kappa \over z} \right) 
\right]
\np
\delta_{s_i  1}  \delta_{s_j  1} \ \delta_{r_i  1} \delta_{r_j  1}
\
\left[
\left( { l   \over   y} - {\kappa \over z} \right) 
\left( {-k^* \over 1 - x} - {\kappa^* \over z} \right)
+
\left( {k^* \over x} - {\kappa^* \over z} \right) 
\left( {-l \over 1 - y} - {\kappa \over z} \right) 
\right]
\np
\delta_{s_i -1}  \delta_{s_j -1} \ \delta_{r_i -1} \delta_{r_j -1}  
\
\left[
\left( { k   \over   x} - {\kappa \over z} \right) 
\left( {-l^* \over 1 - y} - {\kappa^* \over z} \right) 
+
\left( {l^* \over y} - {\kappa^* \over z} \right) 
\left( {-k \over 1 - x} - {\kappa \over z} \right) 
\right]
\np
\delta_{s_i -1}  \delta_{s_j -1} \ \delta_{r_i  1} \delta_{r_j  1} 
\
\left[
\left( { k   \over   x} - {\kappa \over z} \right) 
\left( {-k^* \over 1 - x} - {\kappa^* \over z} \right)
+
\left( {l^* \over y} - {\kappa^* \over z} \right) 
\left( {-l \over 1 - y} - {\kappa \over z} \right) 
\right]
\np
\delta_{s_i  1}  \delta_{s_j  1} \ \delta_{r_i -1} \delta_{r_j  1}
\
\left( { l   \over   y} - {\kappa \over z} \right) 
\, {m_p z \over (1 - y)(1 - x) } \,
\np
\delta_{s_j  1}  \delta_{s_i  1} \ \delta_{r_i  1} \delta_{r_j -1} 
\
\left( {k^* \over x} - {\kappa^* \over z} \right) 
\, {- m_p z \over (1 - y)(1 - x) } \,
\np
\delta_{s_i -1}  \delta_{s_j -1} \ \delta_{r_i -1} \delta_{r_j  1} 
\left( { k   \over   x} - {\kappa \over z} \right) 
\, {m_p z \over (1 - y)(1 - x) } \,
\np
\delta_{s_i -1}  \delta_{s_j -1} \ \delta_{r_i  1} \delta_{r_j -1}     
\left( {l^* \over y} - {\kappa^* \over z} \right) 
\, {- m_p z \over (1 - y)(1 - x) } \,
\np
\delta_{s_i  1}  \delta_{s_j -1} \ \delta_{r_i -1} \delta_{r_j -1}    
\, { m_l z   \over  xy} \,
\left( {-l^* \over 1 - y} - {\kappa^* \over z} \right) 
\np
\delta_{s_j  1}  \delta_{s_i -1} \ \delta_{r_j  1} \delta_{r_i  1}
\, { m_l z   \over  xy} \,
\left( {-k^* \over 1 - x} - {\kappa^* \over z} \right)
\np 
\delta_{s_i -1}  \delta_{s_j  1} \ \delta_{r_i  1} \delta_{r_j  1}    
\, { m_l z   \over  xy} \,
\left( {l \over 1 - y} + {\kappa \over z} \right)
\np
\delta_{s_i -1}  \delta_{s_j  1} \ \delta_{r_j -1} \delta_{r_i -1}
\, { m_l z   \over  xy} \,
\left( {k \over 1 - x} + {\kappa \over z} \right)  
\np
\left( 
\delta_{s_i  1}  \delta_{s_j -1} \ \delta_{r_i -1} \delta_{r_j  1} 
+ 
\delta_{s_i -1}  \delta_{s_j  1} \ \delta_{r_i  1} \delta_{r_j -1} 
\right)
{ m_l m_p z^2  \over  x(1 - x) \, y(1 - y) } \, .
\eeq
One can obtain this result using free spinors with 
physical fermion masses in the fermion currents and 
contracting the Lorentz indices of the currents with 
the indices of the tensor that results from the sum 
over polarizations of the transverse photons exchanged 
between fermions. Therefore, this result can be arrived 
at using the canonical QFT irrespective of the difference 
between the bare and effective particles.

Since the same type of analysis applies to all 
terms in $Y$, we explicitly consider only the 
first spin amplitude
\beq
A_1 \es \delta_{s_i  1}  \delta_{s_j  1} \ \delta_{r_i -1} \delta_{r_j -1} B_1
\, ,
\eeq
where
\beq
B_1
\es
\left( { l   \over   y} - {\kappa \over z} \right) 
\left( {-l^* \over 1 - y} - {\kappa^* \over z} \right) 
+
\left( {k^* \over x} - {\kappa^* \over z} \right) 
\left( {-k \over 1 - x} - {\kappa \over z} \right)  
\\
\es
{ 2|\kappa|^2 \over z^2 }
+
{x + y - 1 \over z } \ 
\left[ 
{|k|^2 \over x(1-x) }
-
{|l|^2 \over y(1-y) } 
\right]
- 
{lk^* \over xy } - { l^*k \over (1-x)(1-y) } \, .
\eeq
The first term in $B_1$ behaves as $1/z^2$
for small $z$, the second term as $1/z$, and the
third term is regular. Only the terms that do not 
change fermion spins involve $1/z^2$. The most 
singular for small $z$ terms appear only in the 
first four spin amplitudes in $Y_+$. Therefore,
they provide the leading behavior of the entire 
$Y_+$ for small values of $z$,
\beq
\label{Yplus}
P^{+\,2} \ Y_+
\es
\delta_{s_i  s_j} \ \delta_{r_i r_j} 
\ 2 (k^\perp - l^\perp)^2 /z^2 \, .
\eeq
It is these leading terms that count most in the 
formation of lepton-proton bound states for small 
electric charge $e$ in the presence of the RGPEP 
form factors in Eqs.~(\ref{cFint}) and (\ref{cFcor}). 
The reason is that the form factors are order 1 
when the invariant mass differences in denominators 
in Eqs.~(\ref{cFint}) and (\ref{cFcor}) are small 
and the same form factors are exponentially small 
when the changes of the invariant masses of fermion 
states are large. The lepton momentum fractions $x$ 
and $y$ must be close to the ratio of the lepton 
mass to the sum of lepton and proton masses and 
cannot change much because this would create a large 
change in the invariant mass and thus also the 
large differences in the denominators and form
factors' arguments. On the other hand, photons are 
massless and can carry small fractions $z$ if their 
$q^\perp$s are sufficiently small not to cause large 
invariant mass changes. This is what happens in the 
atomic-like bound states.

The evaluation of $Y_-$ in Eq.~(\ref{tildeYij1})
yields the same result as Eq.~(\ref{Yplus}).
Therefore, Eqs.~(\ref{AYij}) and (\ref{tildeYij1})
together yield
\beq
\label{AppYij}
Y_{ij}
\es
\sqrt{ \Delta(p_i) \Delta(p_j) \Delta(l_i) \Delta(l_j)} 
\ \Delta(q) \ 
\sqrt{ 4 p_i^+ p_j^+  l_i^+ l_j^+} \
\tilde \delta \  
{ \delta_{s_i  s_j} \ \delta_{r_i r_j} \ 2 (k^\perp - l^\perp)^2 \over q^{+ \, 3} } \, .
\eeq
In the limit of removing the regularization,
\beq
\label{AYijResult1} 
Y_{ij}
\es
\sqrt{ 4 p_i^+ p_j^+  l_i^+ l_j^+} \
\tilde \delta \  
{ \delta_{s_i  s_j} \ \delta_{r_i r_j} \ 2(k^\perp - l^\perp)^2 \over q^{+ \,
    3} } 
\, .
\eeq
Writing $l_i+= xP^+$ and $l_j^+ = y P^+$,
one arrives at
\beq
\label{AYijResult} 
Y_{ij}
\es
\sqrt{ x(1-x) y(1-y)} \ P^{+\,2} \
\tilde \delta  \  
{ \delta_{s_i  s_j} \ \delta_{r_i r_j} \ 4(k^\perp - l^\perp)^2 \over q^{+ \,
    3} } 
\, .
\eeq

\subsection{ Evaluation of $X_{ij}$ }
\label{evXij}

In evaluating the matrix elements of $\hat
X_{4nlA}$, one has to take into account that 
one creation operator for one kind of fermions 
has to be commuted through one creation
operator of the other kind of fermions, with 
net result of a change of sign. The matrix 
element $X_{ij}$ includes the sum of a term 
with $n=l$ and $l=p$ and a term with $n=p$ 
and $l=n$. In the case of $l_i^+ > l_j^+$,
which corresponds to evaluation of $Y_+$ in
App.~\ref{evXij}, one has to consider only terms
that result from the absorption of a photon by 
the lepton, 
\beq
X_{ij}
\es
(-1)
\sqrt{ \Delta(p_i)\Delta(p_j)\Delta(l_i)\Delta(l_j)} \
\Delta(q) \
\sqrt{ p_i^+ p_j^+ l_i^+ l_j^+} \
\tilde \delta \
{ -4 \delta_{s_i  s_j} \delta_{r_i r_j} \
\over q^{+ \, 2} } \ .
\eeq
The result for $Y_-$ is the same. 
After removal of regularization one has
\beq
\label{XijA1} 
X_{ij}
\es
\sqrt{ p_i^+ p_j^+ l_i^+ l_j^+} \
\tilde \delta \
{ 4 \delta_{s_i  s_j} \delta_{r_i r_j} \
\over q^{+ \, 2} } \ .
\eeq
Using the same parameterization of
$l_i^+$ and $l_j^+$ as in Eq.~(\ref{AYijResult}),
one obtains 
\beq
\label{XijA} 
X_{ij}
\es
\sqrt{ x(1-x) y(1-y)} \ P^{+\,2} \
\tilde \delta \
{ 4 \delta_{s_i  s_j} \delta_{r_i r_j} \
\over q^{+ \, 2} } \ .
\eeq

\subsection{ Evaluation of $\cF_{ij \, \rm int}$ and $\cF_{ij \, \rm cor}$ }
\label{Aff}

The denominators to consider are  $i3=-3i$ and $j3=-3j$. 
In both diagrams in Fig.~\ref{Exchange} the calculation 
is carried out in the same fashion. In the case of the 
left diagram,
\beq
p_{i3}/i3 \es 
- q^+ \,
\left[ \kappa^{\perp \, 2}  
     + {m_l^2 z^2 \over x y} 
     + \left( {l^{\perp \, 2} \over y} - {k^{\perp \, 2} \over x} \right) z
\right]^{-1} \, , \\
p_{j3}/j3 \es 
-q^+
\left[ \kappa^{\perp \, 2}  
     + {m_p^2 z^2 \over (1-x)(1-y)} 
     + \left( {k^{\perp \, 2} \over 1 - x} - {l^{\perp \, 2} \over 1-y } \right)z
\right]^{-1} \, .
\eeq
In the square brackets, the first two terms dominate 
for small $z$ and transverse momenta much smaller 
than the fermion masses. These conditions are secured 
in the case of small mass eigenstates of the full 
eigenvalue problem by the presence of RGPEP vertex 
form factors $f$ with large effective particle size 
$s$. The lepton momentum fractions $x$ and $y$ in the 
effective lepton-proton Fock sector are approximately 
\beq
\label{beta}
\beta \es m_l/(m_l + m_p) \, ,
\eeq
and in the leading approximation for transverse momenta 
much smaller than masses and for $x$ and $y$ close to 
$\beta$, one has
\beq
\label{pi3pj3}
p_{i3}/i3 \es
p_{j3}/j3 \rs 
-q^+
\left[ (k^\perp - l^\perp)^2  
     + (m_l+ m_p)^2 z^2 \right]^{-1} \, .
\eeq
This result says that $i3$ and $j3$ for finite
fermion momenta $p_{i3}$ and $p_{j3}$, behave 
as the photon $q^-$ on mass shell plus a correction
due to the $+$ momentum carried by the photon. 
Thus, these denominators diverge when $z \to 0$ 
for fixed $|k^\perp - l^\perp|$.

At the same time, 
\beq
ij \es   
{|k|^2 \over x(1-x) }
-
{|l|^2 \over y(1-y) } 
+
{ m_l^2 \over x } + { m_p^2 \over 1-x }
-
{ m_l^2 \over y } - { m_p^2 \over 1-y } \, ,
\eeq
which is finite for small $z$ and vanishes 
for vanishing $z$ and $|k^\perp - l^\perp|$. 
Therefore, one can neglect $ij$ in comparison 
to $i3$ and $j3$. Consequently, $\cF_{ij \, 
\rm int}$ in Eq.~(\ref{cFint}) reduces to
\beq
\label{cFintA}
\cF_{ij \, \rm int}
\es
-q^+
\left[ (k^\perp - l^\perp)^2  
     + (m_l+ m_p)^2 z^2 \right]^{-1} \, .
\eeq
This result completes calculation of $\cF_{ij \,
\rm int}$.

Evaluation of $\cF_{ij \, \rm cor}$ in 
Eq.~(\ref{cFcor}) now only requires 
estimates of $f_{i3}$ and $f_{3j}$ in
\beq
\label{AcFcor}
\cF_{ij \, \rm cor}
\es
( 1 - f_{ij})
\ f_{i3}f_{3j} \  
{ -q^+
  \over
 (k^\perp - l^\perp)^2  
     + (m_l+ m_p)^2 z^2 } \, ,
\eeq
where $(p_{i3}/i3 + p_{3j}/j3)/2$ is approximated 
using Eq.~(\ref{pi3pj3}). The same approximation 
yields the arguments $i3$ and $j3$ of $f$ in the form 
\beq
\label{i3A}
i3 \es {\beta \over z} \ 
\left[(k^\perp - l^\perp)^2 + (m_l+ m_p)^2 z^2 \right] \ , \\
\label{j3A}
j3 \es {1 - \beta \over z} \ 
\left[(k^\perp - l^\perp)^2 + (m_l+ m_p)^2 z^2 \right] \ .
\eeq
Analogous reasoning applies in the case of $Y_-$. 
The point is that the form factors $f_{i3}$ and $f_{3j}$
vanish exponentially fast for small values of $z$, where
the dominant interaction is active. Therefore,   
$\cF_{ij \, \rm cor}$ is neglected in comparison
with $\cF_{ij \, \rm int}$ in Eq.~(\ref{cFij}). 

\section{ Integration in $d_f$ and $d_a$ }
\label{Adf}

Integration in Eq.~(\ref{df}) must be carried 
out numerically, but it is useful to simplify 
the six-dimensional integral. For the $s$-wave 
ground state one has
\beq
\label{dfA1}
d_f
\es
{
2 \int d^3p \int d^3p\,' \ 
\ { 1 \over (1 + p^2)^2 } \
\left[ e^{ - 4\alpha^4 ( p^2 - p\,'^2)^2/a^4 } - 1 \right] \
{4 \pi \over (\vec p - \vec p \,')^2 }  
\ { 1 \over (1 + p\,'^2)^2 } 
\over
(2\pi)^3
\int d^3p 
\ { 1 \over (1 + p^2)^4 } }
\ .
\eeq
The $z$-axis in integration over $\vec p\,'$
can be chosen along $\vec p$. Integration over 
angles of $\vec p\,'$ replaces $1/\vec q\,^2$ by
\beq
{\pi \over p p'} \ \ln { (p+p')^2 \over (p-p')^2} \ .
\eeq 
There is no dependence left on angles of $\vec p$.
So, integration over all angles yields
\beq
\label{dfA3}
d_{f}
\es
{ 2 \cdot 4\pi \ {\pi \cdot 4\pi}
\over (2\pi)^3 4\pi \cdot \pi/32}
\
\int_0^\infty p dp \int_0^\infty  p' dp' \ 
{ e^{ - 4\alpha^4 ( p^2 - p\,'^2)^2/a^4 } - 1  \over (1 + p^2)^2(1 + p'\,^2)^2 } \
\ \ln{ (p+p')^2\over (p-p')^2}  
\ .
\eeq
Introducing variables $u=p^2$ and $v=p'\,^2$,
one obtains
\beq
\label{dfA4}
d_{f}
\es
{32 \over \pi^2} 
\
\int_0^\infty du \int_0^u  dv \ 
\ { e^{ - z^2 ( u - v )^2} - 1  \over (1 + u)^2(1 + v)^2 } \
\ \ln { \sqrt{u} + \sqrt{v} \over \sqrt{u} - \sqrt{v} }   
\ ,
\eeq
where 
\beq
z^2 \es 
2(\alpha/a)^2 \ .
\eeq
If the parameter $a$ changes from $1/20$ to 3,
the parameter $z$ changes between $8 \ 10^2 \alpha^2$
and $(2/9) \alpha^2$. Changing variables to $zu$ and
$zv$, one has
\beq
\label{dfA5}
d_{f \, S1}
\es
{ 32 \over \pi^2} \ z^2
\
\int_0^\infty du \int_0^u  dv \ 
\ { e^{ - ( u - v )^2} - 1  \over (z + u)^2(z + v)^2 } \
\ \ln { \sqrt{u} + \sqrt{v} \over \sqrt{u} - \sqrt{v} } 
\ . 
\eeq

\end{appendix}



\end{document}